\documentclass[iop]{emulateapj}

\usepackage{natbib}
\usepackage{hyperref}
\usepackage{graphicx}

\newcommand{\msun}{\mbox{M}_{\odot}}
\newcommand{\mstar}{$\mbox{M}_*$}

\newcommand{\secpoint}{\mbox{$''\mskip-7.6mu.\,$}}

\shorttitle{The MOSDEF Survey: Mass, Metallicity, and Star-Formation Rate at $z\sim2.3$}
\shortauthors{Sanders et al.}

\begin{document}

\title{The MOSDEF Survey: Mass, Metallicity, and Star-Formation Rate at $z\sim2.3$\altaffilmark{*}}
\altaffiltext{*}{Based on data obtained at the
W.M. Keck Observatory, which is operated as a scientific partnership among the California Institute of Technology, the
University of California, and NASA, and was made possible by the generous financial support of the W.M. Keck Foundation.
}

\author{Ryan L. Sanders\altaffilmark{1}} \altaffiltext{1}{Department of Physics \& Astronomy, University of California, Los Angeles, 430 Portola Plaza, Los Angeles, CA 90095, USA}

\author{Alice E. Shapley\altaffilmark{1}}

\author{Mariska Kriek\altaffilmark{2}} \altaffiltext{2}{Astronomy Department, University of California, Berkeley, CA 94720, USA}

\author{Naveen A. Reddy\altaffilmark{3,5}} \altaffiltext{3}{Department of Physics \& Astronomy, University of California, Riverside, 900 University Avenue, Riverside, CA 92521, USA}

\author{William R. Freeman\altaffilmark{3}}

\author{Alison L. Coil\altaffilmark{4}} \altaffiltext{4}{Center for Astrophysics and Space Sciences, University of California, San Diego, 9500 Gilman Dr., La Jolla, CA 92093-0424, USA}

\author{Brian Siana\altaffilmark{3}}

\author{Bahram Mobasher\altaffilmark{3}}

\author{Irene Shivaei\altaffilmark{3}}

\author{Sedona H. Price\altaffilmark{2}}

\author{Laura de Groot\altaffilmark{3}}

\altaffiltext{5}{Alfred P. Sloan Fellow}

\email{email: rlsand@astro.ucla.edu}

\begin{abstract}
We present results on the $z\sim 2.3$ mass-metallicity relation (MZR)
using early observations from the MOSFIRE Deep Evolution
Field (MOSDEF) survey. We use an initial sample of 87 star-forming
galaxies with spectroscopic coverage of H$\beta$, [O~\textsc{iii}]$\lambda 5007$,
H$\alpha$, and [N~\textsc{ii}]$\lambda 6584$ rest-frame optical emission
lines, and estimate the gas-phase oxygen abundance based on the N2
and O3N2 strong-line indicators. We find a positive correlation between
stellar mass and metallicity among individual $z\sim 2.3$ galaxies using
both the N2 and O3N2 indicators. We also measure the emission-line
ratios and corresponding oxygen abundances for composite spectra
in bins of stellar mass. Among composite spectra, we find a monotonic
increase in metallicity with increasing stellar mass, offset
$\sim 0.15 - 0.3$ dex below the local MZR. When the sample is divided
at the median star-formation rate (SFR), we do not observe significant SFR
dependence of the $z\sim 2.3$ MZR among either individual galaxies
or composite spectra. We furthermore find that $z\sim 2.3$ galaxies have metallicities $\sim0.1$
dex lower at a given stellar mass and SFR than is observed locally.
 This offset suggests that high-redshift galaxies do not fall on the local
``fundamental metallicity relation" among stellar mass, metallicity,
and SFR, and may provide evidence of a phase of galaxy
growth in which the gas reservoir is built up due to
inflow rates that are higher than star-formation and outflow rates.
However, robust conclusions regarding the gas-phase oxygen abundances
of high-redshift galaxies await a systematic reappraisal of the application
of locally calibrated metallicity indicators at high redshift.
\end{abstract}


\section{Introduction}\label{sec:intro}

The study of chemical abundances in galaxies at various epochs in cosmic history highlights key processes
 governing the growth and evolution of galaxies.  In the local universe, there is a clear relationship between the
 stellar mass (\mstar) of a galaxy and its gas-phase oxygen abundance,
 such that galaxies with lower stellar masses have lower metallicities than those with higher stellar masses.
  The $z\sim0$ mass-metallicity relationship (MZR) has been confirmed by many studies \citep[e.g.,][]{tre04,kew08,and13}.
  Local galaxies follow this relationship with an intrinsic scatter of $\sim0.1$ dex.
  The MZR has been confirmed at redshifts up to $z\sim3.5$ and has been observed to evolve with redshift, such that
 galaxies of a given stellar mass have lower metallicities at higher redshifts \citep[e.g.,][]{erb06, mai08,hen13,mai14,mas14,ste14}.
  The MZR is most commonly understood in terms of the interplay between star formation and gas flows.
  As the stellar content of a galaxy grows over time, the chemical enrichment in the ISM increases due to the recycling of
 heavy elements produced in stars back into the ISM.
  This process of pure enrichment is modulated by gas inflows and outflows \citep[e.g.,][]{fin08,man10,dav11,dav12} which may either
 increase or decrease the enrichment depending on the metallicity of the gas flow.

  Much insight can be gained from the form and evolution of the MZR if the details of the underlying physical
 processes are understood.  It has been suggested that the  MZR arises from the interaction of a galactic wind
 with the gravitational potential of a galaxy \citep{dek86,tre04}.  In this scenario, less massive galaxies are
 naturally less enriched as it is easier for winds to escape the gravitational potential well and remove metals in the
 process.  At high stellar masses, winds are unable to escape
 and the galaxy retains all of the heavy elements injected into the ISM, naturally explaining the asymptotic behavior of the MZR
 assuming a constant stellar yield.  Alternatively, in the equilibrium model of \citet{fin08}
 and \citet{dav11,dav12}, outflows remove some metals,
 but have a more important effect of decreasing the fraction of inflowing gas from the intergalactic medium
 that is able to form stars and produce
 metals.  In these models, the mass-loading factor quantifies the efficiency with which winds remove material from galaxies.
  Since the mass-loading factor of the momentum-driven winds in the equilibrium model scales
 inversely with \mstar, the star-formation efficiency
 in low-\mstar~galaxies is drastically lowered and fewer metals are produced.
  In this context, the low-mass slope of the MZR can probe how the mass-loading factor scales with galaxy mass \citep{fin08}.
  Other explanations attribute the MZR to variations in the
 star formation efficiency \citep{tas08} or the gas mass fraction \citep{zah14} without referencing gas flows, although these
 processes are undoubtedly linked.

The local MZR has been observed to have a significant star-formation rate (SFR) dependence.  \citet{man10} found that
 local star-forming galaxies lie on a two-dimensional surface in \mstar-SFR-metallicity space with a scatter of
 only 0.05 dex in metallicity.  This surface is a strong function of SFR at low stellar masses such that galaxies of a given
 \mstar~with higher SFRs have lower metallicities, while only showing weak SFR dependence at high stellar masses.
  This relationship among \mstar, SFR, and metallicity is referred to as the ``fundamental metallicity relation" (FMR), and
 the MZR is a projection of the FMR in the \mstar-metallicity plane.
  The existence of a local FMR has been confirmed by recent studies \citep{and13,lar13}.
  The connection between SFR and metallicity has been interpreted as a signature of infalling pristine gas which dilutes
 the metals in the ISM while simultaneously fueling additional star formation.  If there is no inflow, the ISM enrichment
 increases while star formation naturally decreases as gas is used up.
  \citet{man10} find that high-redshift galaxies fall on the same FMR as local galaxies,
 naturally explaining redshift evolution in the MZR as a result of the SFR at fixed \mstar~increasing with redshift.
  More recently, \citet{lil13} showed that a non-evolving FMR is a natural consequence of a physical model of galaxies
 in which the SFR is regulated by the mass of the gas reservoir if the dependence of the gas depletion timescale and
 the mass-loading factor on stellar mass is constant with redshift.
  However, the existence of the FMR has not been confirmed
 at redshifts above $z\sim1$ as large and unbiased samples have been difficult to obtain up to this point and inconsistencies
 between different metallicity indicators and calibrations make comparisons difficult.  Whether or not
 high-redshift galaxies lie on an extension of the local FMR, or follow an FMR at all, remains controversial
 \citep{bel13,sto13,cul14,ste14,tro14}.

In this work, we present early observations from the MOSFIRE Deep Evolution Field (MOSDEF) survey, which will
 contain rest-frame optical spectra of $\sim$1500 galaxies at $z\sim1.5-3.5$ when completed.
  Here, we focus on an initial sample of 87 star-forming galaxies
 at $z\sim2.3$ with estimates of stellar masses, gas-phase oxygen abundances, and H$\alpha$-based dust-corrected SFRs.
  We study the MZR at $z\sim2.3$ with a representative sample of individual high-redshift measurements
 with a wide wavelength coverage and a large dynamic range in stellar mass and [N~\textsc{ii}]/H$\alpha$ ratio.  While this initial
 MOSDEF sample already surpasses nearly all samples in the literature used to study the MZR at this redshift, the full sample
 will provide a much larger dataset than has previously been available.  In Section~\ref{sec:data}, we introduce and give
 a brief overview of the MOSDEF survey and describe our observations, measurements, and sample selection.  In
 Section~\ref{sec:met}, we describe the metallicity calibrations, present the $z\sim2.3$ MZR, and investigate
 dependence on SFR.  Finally, we summarize and discuss our results in Section~\ref{sec:disc}.  We assume a standard
 $\Lambda$CDM cosmology throughout with $\Omega_m=0.3$, $\Omega_{\Lambda}=0.7$, and H$_0=70$ km s$^{-1}$ Mpc$^{-1}$.
  Throughout this paper, the term metallicity refers to the gas-phase oxygen abundance ($12+\log(\mbox{O/H})$), which acts as a
 proxy for the true gas-phase metallicity.

\section{Data}\label{sec:data}

\subsection{The MOSDEF Survey} \label{sect:data-mosdef}
The MOSDEF survey is a four-year
project using the MOSFIRE spectrograph \citep{mcl12} on
the 10~m Keck~I telescope to survey the physical properties
of galaxies at $1.4 \leq z \leq 3.8$.
The full details of the survey
will be presented in Kriek et al. (in prep.), but here we describe
its basic parameters.
With MOSDEF, we target galaxies in the regions of the AEGIS, COSMOS, and GOODS-N
extragalactic legacy fields with {\it Hubble Space Telescope} ({\it HST}) imaging
coverage from the CANDELS survey \citep{gro11,koe11}, totaling
500 square arcminutes.  All MOSDEF targets have extensive multi-wavelength ancillary data
including {\it Chandra}, {\it Spitzer}, {\it Herschel}, {\it HST}, VLA, and
ground-based optical and near-IR observations.
The majority of this area is also covered 
by the 3D-HST grism survey \citep{bra12a}.

When complete, the MOSDEF survey will consist of rest-frame optical
spectra for $\sim 1500$ galaxies in three distinct redshift intervals
($\sim 750$~galaxies at $2.09 \leq z \leq 2.61$, $\sim 400$
at $1.37 \leq z \leq 1.70$, and $\sim 400$ at $2.95 \leq z \leq 3.80$).
Each range is dictated by the redshifts at which strong rest-frame
optical emission features fall within windows of atmospheric transmission.
Based on photometric catalogs compiled by the 3D-HST team \citep{ske14},
galaxies are targeted down to limiting {\it HST}/WFC3 F160W magnitudes of 24.0,
24.5, and 25.0, respectively, at $z\sim 1.5, 2.3$, and $3.4$. Target priorities
are determined by both apparent brightness and existing redshift information,
according to which brighter galaxies and those with more secure redshift
information are assigned higher priority. We adopt photometric and
grism redshifts from the 3D-HST survey, while 
additional redshift information is assembled in the form of
ground-based spectroscopic redshifts from various sources. We note that only $\sim 40$\% of MOSDEF
targets observed to date had prior spectroscopic redshifts.

\subsection{Observations, Data Reduction, and Measurements} \label{sect:data-observations}
As described in Kriek et al. (in prep),
MOSDEF observations are designed to maximize the number of strong
rest-frame optical spectroscopic features covered at $3700 - 7000$~\AA\
that are accessible from the ground. In this paper, we focus on the 
$z\sim 2.3$ redshift interval, for which we obtain J, H, and K-band spectra.
The nominal wavelength coverage in each of these bands is 
$1.153-1.352 \mu\mbox{m}$ (J-band), $1.468-1.804 \mu\mbox{m}$ (H-band),
and $1.954-2.397 \mu\mbox{m}$ (K-band), but varies slightly depending
on the horizontal slit location in the mask. The average exposure time for each mask is
2 hours per filter, reaching unobscured SFRs of 
$\sim 1 \mbox{ M}_{\odot} \mbox{ yr}^{-1}$ as traced by Balmer emission lines.

The data presented here are drawn from the first observing season of the MOSDEF
survey, spanning five observing runs from 2012 December to 2013 May and comprising eight
MOSFIRE masks.\footnote[6]{In addition to data collected on observing runs specifically scheduled
for the MOSDEF project, H- and K-band observations were obtained by K. Kulas, I. McLean,
and G. Mace in 2013 May for one MOSDEF mask in the GOODS-N field.}
Due to the range of field visibility,
two of the masks observed during a pilot run in 2012 December
target additional CANDELS legacy fields: one mask in GOODS-S and one in UDS.
Each mask typically had $\sim 30$ 0\secpoint7 slits, yielding a resolution of, respectively,
3300 in J, 3650 in H, and 3600 in K. 
As motivated in Kriek et al. (in prep.), masks were generally observed using an ABA'B'
dither sequence with 1\secpoint5 and 1\secpoint2 outer and inner nod amplitudes. However,
during the first MOSDEF run in 2012 December, we experimented with an ABBA dither pattern.
Individual exposure times within a dither sequence consisted of 180 seconds in
K, and 120 seconds in J and H. The seeing as measured in 
individual exposures ranged from 0\secpoint35 to 1\secpoint65,
with a median value of 0\secpoint65, and photometric conditions ranged from clear to
variable. For galaxies at $z\sim2.3$, 
the strongest features of interest are [O~\textsc{ii}]$\lambda\lambda$3726,3729
in the J band, H$\beta$ and [O~\textsc{iii}]$\lambda\lambda$4959,5007 in the H band,
and H$\alpha$, [N~\textsc{ii}]$\lambda$6584, and [S~\textsc{ii}]$\lambda\lambda$6717,6731
in the K band. Specifically, we use combinations of  H$\beta$, [O~\textsc{iii}],
H$\alpha$, and [N~\textsc{ii}] emission-line fluxes for metallicity estimates,
H$\alpha$/H$\beta$ line ratios for estimates of dust extinction, and dust-corrected
H$\alpha$ luminosities for estimates of SFRs.

The data were reduced using a custom IDL pipeline (see Kriek et al., 
in prep., for a full description). In brief, science frames
were cut up into individual two-dimensional slits,
flatfielded, sky-subtracted, wavelength-calibrated, cleaned of cosmic rays,
rectified, combined, and flux-calibrated. The relative spectral response was
estimated using observations of B8 -- A1 V standard stars matched in air mass
to the science observations, while an initial absolute calibration was achieved
by forcing the flux density in the spectrum of a reference star on the mask to match its cataloged
photometry.
  Flux densities between different filters were verified to be consistent for those objects
 with detected continuum, confirming that the absolute calibration is valid across all filters.
 Two-dimensional error spectra were calculated including the
effects of both Poisson counts from the sky and source as well as read noise.
One-dimensional science and error spectra for the primary target in each slit were
then optimally extracted \citep{hor86}, along with those of any serendipitous objects detected
(Freeman et al., in prep.).
The initial absolute flux calibration for each spectrum was refined by estimating
the amount of slit loss for each target relative to that for the reference star --
a function of a two-dimensional elliptical Gaussian fit to the {\it HST} F160W
image of the galaxy convolved with the average seeing profile estimated for each mask and filter.
  After slit loss correction, the flux densities of objects with detected continuum were
 consistent with broadband photometric measurements, and the systematic offset between spectroscopic
 and photometric flux densities was small compared to other sources of uncertainty.

Emission-line fluxes were measured by fitting Gaussian line profiles to the extracted, flux-calibrated one-dimensional spectra.
Uncertainties on the emission-line fluxes were estimated by perturbing the one-dimensional spectrum many times according to the error
spectrum, refitting the line profile, and measuring the width of the resulting
distribution of fluxes.  Redshifts were measured from the observed centroids of the highest signal-to-noise (S/N) emission
lines, typically H$\alpha$ or [O~\textsc{iii}]$\lambda$5007.

Stellar masses were estimated using the MOSDEF redshifts and pre-existing photometric data assembled by the 3D-HST team \citep{ske14}.
We modeled the photometric dataset for each galaxy with the SED-fitting program FAST \citep{kre09}, assuming the stellar population 
synthesis models of \citet{con09} and a \citet{cha03} IMF. Star-formation histories were parameterized using so-called delayed-$\tau$ 
models of the form SFR($t$)$\propto t e^{-t/\tau}$, where $t$ is the stellar-population age, and $\tau$ is the e-folding timescale in the SFR.
Dust extinction was described using the \citet{cal00} attenuation curve.
For each galaxy, a grid in stellar population age, e-folding timescale, metallicity, and dust extinction was explored
and $\chi^2$ minimization was used to find the best-fitting model. The normalization of the best-fit model yielded
the stellar mass. Confidence intervals in each stellar population parameter were determined using Monte Carlo simulations
where the input SED was perturbed and refit 500 times.
SFRs are based on dust-corrected H$\alpha$ luminosites.  Dust corrections are estimated from the ratio of H$\alpha$/H$\beta$,
 in which H$\alpha$ and H$\beta$ fluxes have been corrected for underlying Balmer absorption (Reddy et al., in prep.).
  Balmer absorption equivalent widths for H$\alpha$ and H$\beta$ are measured from the best-fit SED model for each galaxy.
  \textit{E(B-V)$_{neb}$} is
 calculated assuming an intrinsic ratio of 2.86 and using the dust-attenuation curve of \citet{cal00}.
  Dust-corrected H$\alpha$ luminosities are translated into SFRs
 using the calibration of \citet{ken98}, converted to a \citet{cha03} IMF.

\subsection{Sample Selection}

\begin{figure}
 \centering
 \includegraphics[width=\columnwidth]{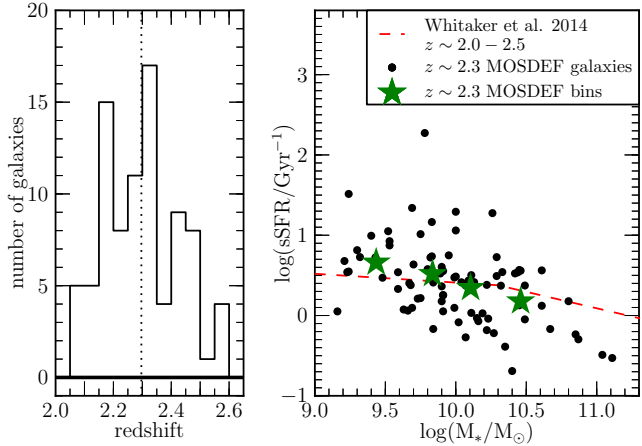}
 \caption{Properties of the $z\sim2.3$ MOSDEF MZR sample, containing 87 star-forming galaxies.
  \textsc{Left:} Redshift distribution,
 in which the vertical dotted line
 shows the mean redshift of $\langle z\rangle=2.296$ with a standard deviation of 0.126.
  \textsc{Right:} sSFR vs. \mstar~for the $z\sim2.3$ sample.  Black points show individual MOSDEF galaxies.
  The red dashed line shows the broken power-law fit to the $2.0<z<2.5$ star-forming sequence from
 \citet{whi14}, with SFRs based on IR and UV luminosity.  Green stars show the median sSFR and \mstar~after dividing the MOSDEF sample into four bins in stellar
 mass such that each bin contains $\sim22$ galaxies.
}\label{fig:samplechar}
\end{figure}

We select a sample of $z\sim2.3$ star-forming galaxies from early MOSDEF observations, requiring the following criteria:
\begin{enumerate}
\item Redshift in the range $2.08\le z\le2.61$ in order to have spectral coverage of H$\alpha$, H$\beta$,
 [O~\textsc{iii}]$\lambda$5007, and [N~\textsc{ii}]$\lambda$6584.
\item $\mbox{S/N}\ge3$ for H$\alpha$ and H$\beta$ to reliably measure the dust-corrected star formation rate.
\item Objects must not be flagged as an active galactic nucleus (AGN) in the MOSDEF catalog, identified by X-ray luminosity and infrared colors \citep{don12}.
  Additionally, we require $\log($[N~\textsc{ii}]$\lambda$6584/H$\alpha)<-0.3$.
\end{enumerate}
These criteria result in a sample of 88 $z\sim2.3$ star-forming galaxies.  One object is excluded from the sample because
 of a lack of wavelength coverage of [O~\textsc{iii}]$\lambda$5007 due to the location of the slit on the mask.  Our final sample is
 therefore 87 galaxies with an average redshift of $\langle z\rangle =2.296\pm0.126$.
  The sample redshift distribution is shown in the left panel of Figure~\ref{fig:samplechar}.
  The right panel of Figure~\ref{fig:samplechar} shows the specific SFR (SFR/M$_*$; sSFR) as a function of
 stellar mass for this sample, where sSFR values are based on dust-corrected H$\alpha$ luminosities.
  The red dashed line shows the broken power-law fit to the star-forming sequence at $2.0<z<2.5$
 by \citet{whi14}, where SFR was determined using IR and UV luminosity.
  Our MOSDEF sample of $z\sim2.3$ star-forming galaxies is consistent with this sequence
 and does not show any obvious bias towards high SFR at a given stellar mass.  To better understand the biases
 of our sample, we divide the sample into four bins in stellar mass such that each bin contains approximately the same
 number of galaxies and take the median sSFR and \mstar~of each bin, shown as green stars.  Our high-redshift sample
 is quite representative across the mass range $\log($\mstar$/\msun)\sim9.4-10.5$, although there may be a very slight
 bias toward high-SFR galaxies at low stellar masses.

In order to study the evolution of the MZR from $z\sim0$ to $z\sim2.3$, we select a sample of local
 galaxies from the Sloan Digital Sky Survey \citep[SDSS;][]{yor00} Data Release 7 \citep[DR7;][]{aba09} catalog.
  Emission-line measurements and galaxy properties are taken from the MPA-JHU catalog of measurements for SDSS DR7.\footnote[7]{Available at \texttt{http://www.mpa-garching.mpg.de/SDSS/DR7/}}
  We require the following criteria:
\begin{enumerate}
\item Minimum redshift restriction of $z\ge0.04$ to avoid aperture effects.
\item Maximum redshift restriction of $z<0.1$ to keep the sample local and avoid any redshift evolution in the MZR,
 observed at redshifts as low as $z\sim0.3$ \citep{lar13}.
\item Measured stellar mass \citep{kau03a}.
\item $\mbox{S/N}\ge5$ for H$\alpha$, H$\beta$, [O~\textsc{iii}]$\lambda$5007, and [N~\textsc{ii}]$\lambda$6584.
\end{enumerate}
AGN are rejected using the criterion of \citet{kau03b}, producing a comparison sample of 70,321 local galaxies.

\section{Metallicity}\label{sec:met}

We use the N2 ($\log{(\mbox{[N~\textsc{ii}]}\lambda\mbox{6584/H}\alpha)}$) and O3N2
 ($\log{((\mbox{[O~\textsc{iii}]$\lambda$5007/H$\beta$})/(\mbox{[N~\textsc{ii}]$\lambda$6584/H$\alpha$}))}$)
 indicators to estimate oxygen abundances.  For both indicators, we use the calibrations of \citet{pet04}
 based on a sample of H~\textsc{ii} regions with direct electron temperature measurements.  These calibrations are given by
\begin{equation}\label{eq:pp04n2}
12+\log{(\mbox{O/H})}=8.90+0.57\times\mbox{N2}
\end{equation}
\begin{equation}\label{eq:pp04o3n2}
12+\log{(\mbox{O/H})}=8.73-0.32\times\mbox{O3N2}
\end{equation}
where $12+\log{(\mbox{O/H})}$ is the oxygen abundance.  The N2 and O3N2 calibrations
 have systematic uncertainties of 0.18 and 0.14 dex, respectively.  Analyses were also performed using the N2 calibration of
 \citet{mai08}, but these results are omitted from this study as they are very similar to those
 based on the \citet{pet04} N2 calibration.

\begin{figure*}
 \centering
 \includegraphics[width=0.495\textwidth]{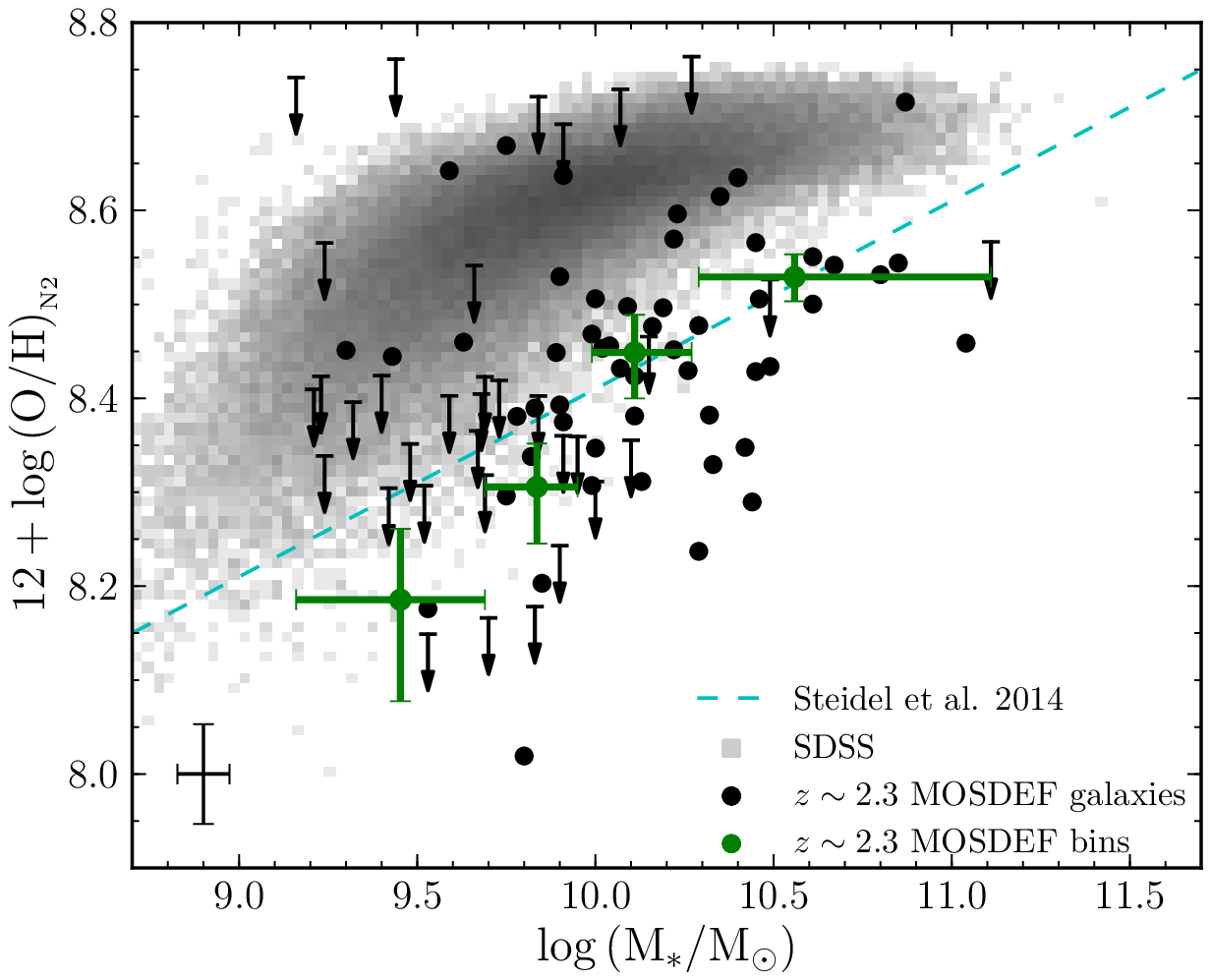}
 \includegraphics[width=0.495\textwidth]{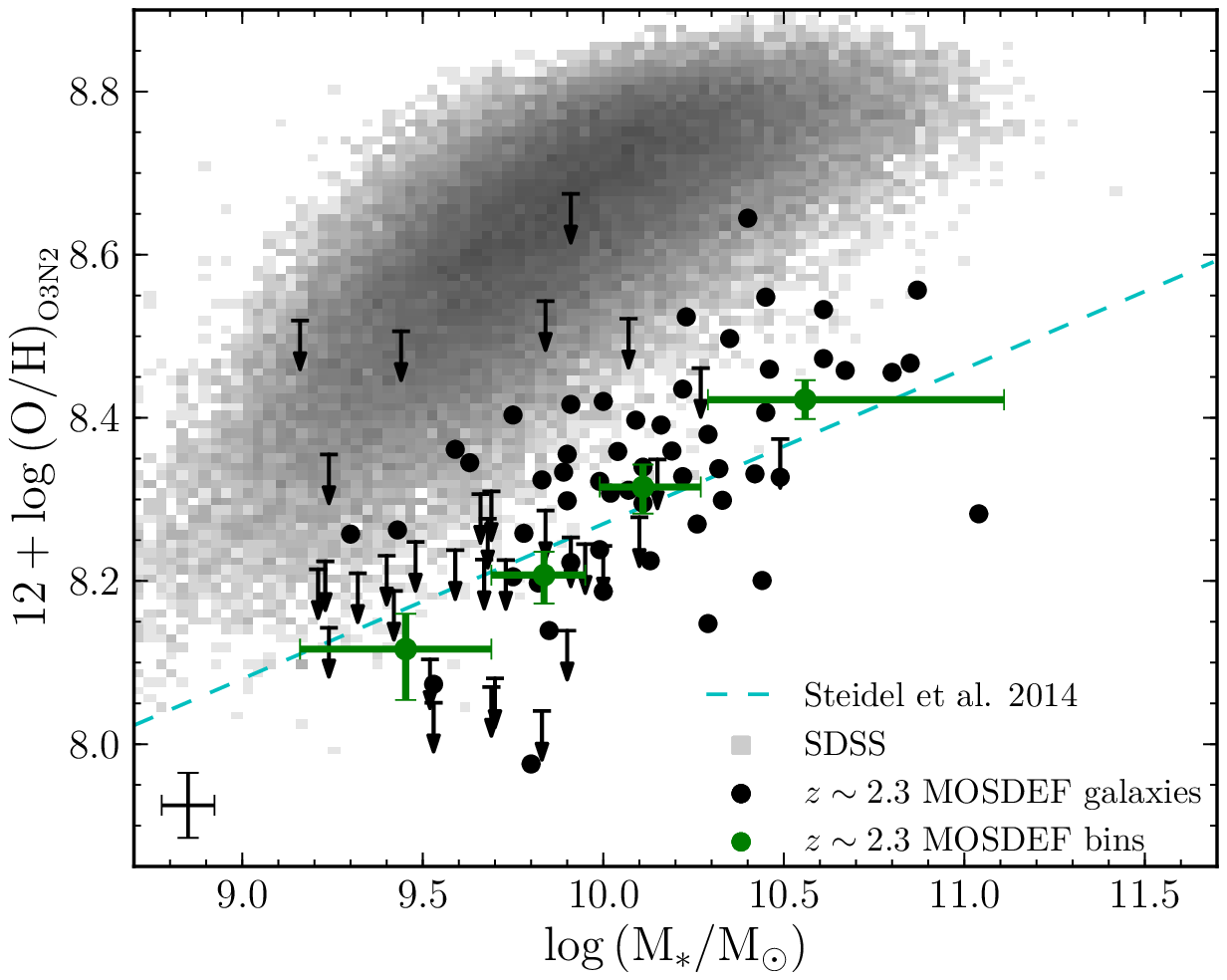}
 \caption{The MZR for $z\sim2.3$ star-forming galaxies with metallicities determined using the
 N2 (left) and O3N2 (right) indicators.  Black points indicate MOSDEF galaxies with 3$\sigma$ or greater significance in H$\alpha$,
 [N~\textsc{ii}]$\lambda$6584, H$\beta$, and [O~\textsc{iii}]$\lambda$5007.  Black arrows indicate 3$\sigma$ upper limits where
 [N~\textsc{ii}]$\lambda$6584 was not detected.  The black error bar in the lower left-hand corner shows the mean uncertainty
 in $12+\log{(\mbox{O/H})}$ and stellar mass for individual MOSDEF galaxies, excluding the calibration error.  The gray
 two-dimensional histogram shows the density of local SDSS galaxies in this parameter space.  Green points with error bars
 represent stacks of individual $z\sim2.3$ galaxies containing $\sim22$ galaxies each, in which the oxygen abundance  was estimated
 from the emission-line measurements of composite spectra and the stellar mass is plotted at the mean $\log($\mstar$/\msun)$.
  The vertical error bar indicates the uncertainty in oxygen
 abundance estimated from the uncertainty on composite emission-line fluxes, excluding the calibration error.  The horizontal
 error bar shows the range of \mstar~for a given bin.  The cyan dashed line shows the best-fit line to the
 $z\sim2.3$ N2 and O3N2 MZR as observed by \citet{ste14}.
}\label{fig:mzr}
\end{figure*}

The MZR for $z\sim2.3$ star-forming galaxies from the MOSDEF sample is shown in
 Figure~\ref{fig:mzr} with metallicities determined using the N2 indicator (left) and O3N2 indicator (right).
  We present 53 individual detections and 34 upper limits: the largest rest-frame optical selected
 sample of individual measurements for which the MZR has been observed at $z>2$.
  Black points indicate $z\sim2.3$ MOSDEF galaxies with detections of all lines, downward arrows indicate 3$\sigma$
 upper limits where [N~\textsc{ii}]$\lambda$6584 was not detected at 3$\sigma$ significance or greater, and
 gray blocks show the density of local SDSS galaxies.  The mean uncertainty on a single MOSDEF point, excluding
 the calibration uncertainty,
 is shown by the black error bar in the lower left-hand corner.
  We observe a progression in metallicity with increasing mass among the individual $z\sim2.3$ galaxies, with the upper
 limits suggesting this trend continues to lower metallicities at low stellar masses.
  The scatter amongst
 individual points is large with respect to the range of parameter space covered, with smaller scatter in the O3N2 MZR than
 in the one based on N2.
  When we include only detections, the
 N2 and O3N2 MZRs have Spearman correlation coefficients of 0.31 and 0.53, respectively, corresponding to
 likelihoods of 0.022 and $4.0\times10^{-5}$ that \mstar~and metallicity are uncorrelated.

There are three detections in the mass range $\log($\mstar$/\msun)\sim9.5-10$ with unusually large
 [N~\textsc{ii}]$\lambda$6584/H$\alpha$ ratios for their stellar masses,
 causing these galaxies to appear as outliers with high metallicities
 in the N2 MZR.  Of these three objects, two also have high [O~\textsc{iii}]$\lambda$5007/H$\beta$ ratios and are offset in the
 [O~\textsc{iii}]/H$\beta$ vs. [N~\textsc{ii}]/H$\alpha$ diagnostic diagram \citep[BPT diagram;][]{bpt81}
 into a region occupied by AGN in the local universe, lying slightly above both the \citet{kew01} $z\sim0$ ``maximum-
starburst" line and the ``evolved" $z\sim2$ line from \citet{kew13t}.  
  Thus, we consider these two objects to be potential optical AGN candidates.
  However, given that high-redshift star-forming galaxies are offset from the local star-forming sequence in the
 BPT diagram \citep[]{sha05,liu08,kew13,sha14,ste14}, local demarcations dividing star-forming galaxies
 and AGN in this parameter space likely need to be revised for application at high redshifts \citep[see][]{coi14}.
  There is no indication based on X-ray properties and
 rest-frame near-IR colors that these two objects are AGN
 \citep{coi14}, although the X-ray upper limits on these objects are not very constraining,
 and furthermore none of the N2 outliers are offset in the O3N2 MZR.  For these reasons, we retain
 these objects in the sample as star-forming galaxies, although removing them from the sample has a negligible effect
 on the results presented in Figure~\ref{fig:mzr}.

The cyan dashed line in Figure~\ref{fig:mzr} shows the best fit to the $z\sim2.3$ N2 and O3N2 MZR as observed
 by \citet{ste14} based on independent datapoints.
  While generally consistent in normalization,
 the MOSDEF sample suggests a steeper slope of the MZR than the sample of \citet{ste14}.
  This inconsistency can be at least partially attributed to a difference in selection criteria.
  The sample of \citet{ste14} is rest-frame ultraviolet selected, which results in biases against galaxies with low SFRs
 and low stellar masses, as well as abundantly dusty galaxies at all masses \citep{red12}.
  The MOSDEF sample is rest-frame optical selected and less susceptible to these biases, as shown in Figure~\ref{fig:samplechar}.
  It is interesting to note that \citet{ste14} observe higher metallicities at low \mstar~where their sample
 has higher typical SFRs than the MOSDEF sample.  This offset is contrary to what one would expect based on the local FMR.
  \citet{ste14} also use a lower detection threshold, considering
 lines with 2$\sigma$ significance as detections, which could have an effect on the observed low-mass slope of the MZR where the
[N~\textsc{ii}]$\lambda$6584 line is very weak.

\begin{figure*}
 \centering
 \includegraphics[width=\textwidth]{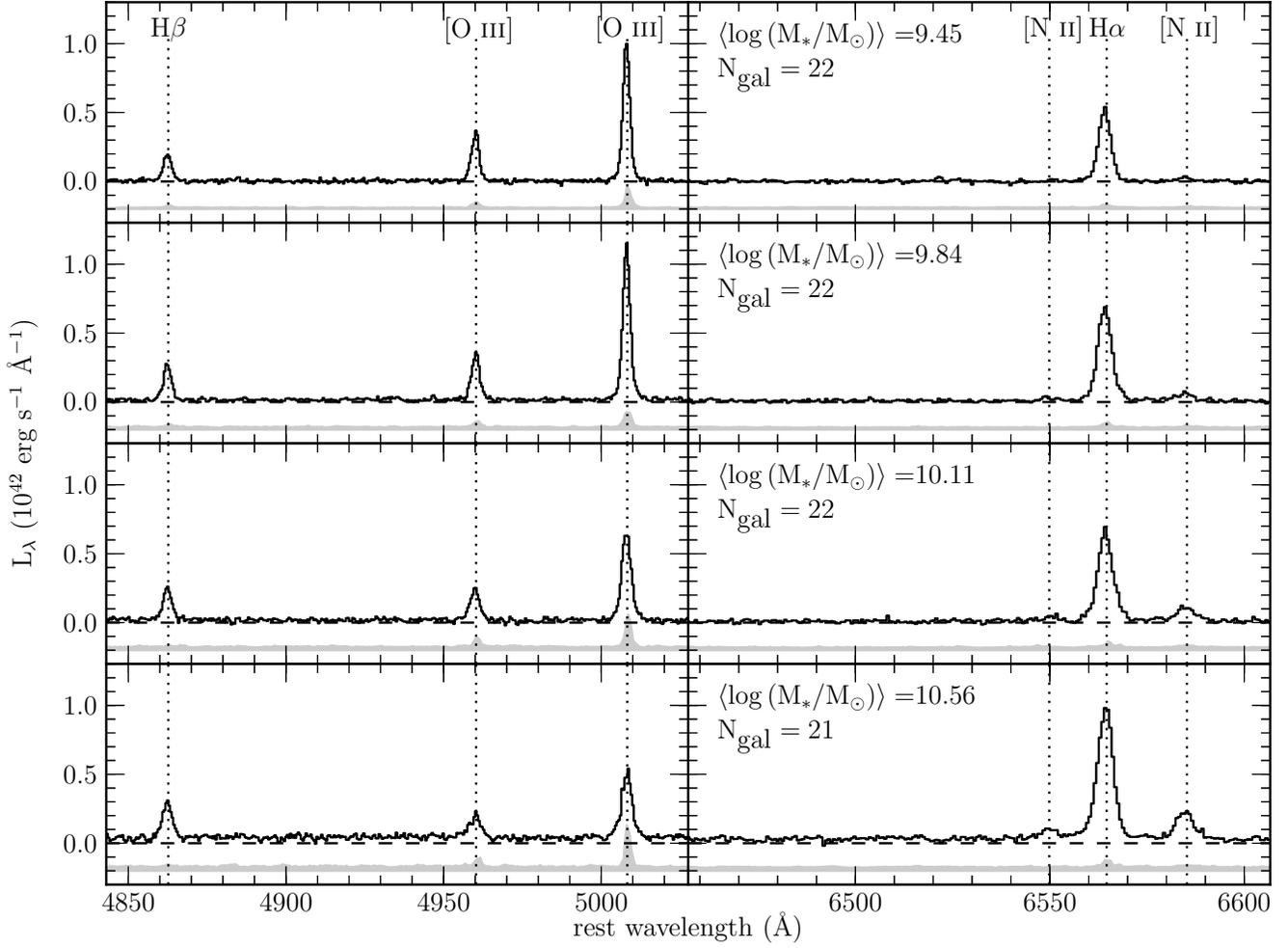}
 \caption{Composite spectra for the $z\sim2.3$ sample separated into four stellar mass bins.  Composite error spectra are
 shown as a gray band offset below the composite spectra for clarity.
  Each bin contains $\sim22$ galaxies.
  The average $\log($\mstar$/\msun)$ for each bin is shown, with mass increasing from top to bottom.  Dotted vertical lines
 highlight strong rest-frame optical emission lines.
  From left to right, these lines are H$\beta$, [O~\textsc{iii}]$\lambda$4959, [O~\textsc{iii}]$\lambda$5007,
 [N~\textsc{ii}]$\lambda$6548, H$\alpha$, and [N~\textsc{ii}]$\lambda$6584.  
}\label{fig:stackspec}
\end{figure*}

To determine more clearly where the $z\sim2.3$ MZR lies with respect to the local MZR, we separated the MOSDEF galaxies
 into four bins in stellar mass such that each mass bin had approximately the same number of galaxies and created
 a composite spectrum
 for each mass bin.  Individual spectra were first shifted into the rest-frame, converted from flux density to luminosity
 density, and normalized by H$\alpha$ luminosity in order to obtain mean line ratios of the galaxies in the bin, as well as
 prevent high-SFR galaxies from dominating the composite spectrum.  The H$\alpha$-normalized spectra were interpolated on
 a grid with wavelength spacing equal to the rest-frame wavelength spacing of the average redshift of the sample.  This
 yields wavelength spacings of 0.49~\AA\ in the H-band and 0.66~\AA\ in the K-band.  At each wavelength increment, the median
 value of the normalized spectra in the bin was selected to create a normalized composite spectrum.  The normalized
 composite spectrum was then multiplied by the average H$\alpha$ luminosity in that bin to give the final composite
 spectrum in units of luminosity density (erg~s$^{-1}$~\AA$^{-1}$).
  In order to create error spectra for the composite spectra, we first perturbed the stellar masses according to their
 uncertainties assuming a log-normal distribution, then separated the objects into four stellar mass bins with the same mass
 ranges used to produce the original composite spectra.  Within each bin, we bootstrap resampled to account
 for sample variance and
 perturbed the spectrum of each object in the bootstrap sample according to the error spectrum of that object to account for
 measurement uncertainty.  The resulting perturbed spectra were combined to form a new composite spectrum.  This process
 was repeated 2500 times to build up a well-sampled distribution of luminosities for each wavelength increment.
  The magnitude of the error spectrum at a given wavelength is half of the 68th-percentile width of this distribution.
  The composite spectra and composite error spectra of the four stellar mass bins for the $z\sim2.3$ sample
 are shown in Figure~\ref{fig:stackspec}.

Emission-line fluxes were measured by fitting Gaussian line profiles to emission lines in the composite spectra.  As can
 be seen in the composite spectra, none of the lines of interest are blended, so lines were fit separately with single
 Gaussian profiles.  We required [N~\textsc{ii}]$\lambda$6584 to have the same width as H$\alpha$.  Uncertainties on the line fluxes
 were estimated
 using the 68th-percentile half-width of the distribution of line fluxes obtained by
 perturbing the composite spectrum according to the composite error spectrum and remeasuring the line fluxes 1000 times.
  Measured emission-line fluxes and uncertainties for the composite spectra were converted
 into metallicities using equations~(\ref{eq:pp04n2}) and~(\ref{eq:pp04o3n2}).
  Emission-line measurements and oxygen abundances from the composite spectra are presented in Table~\ref{tab:binvalues},
 as well as the average galaxy properties of the bins.
  We note that in the lowest stellar mass bin, [N~\textsc{ii}]$\lambda$6584 has a significance lower than 3$\sigma$ with respect to
 the uncertainty on the emission-line flux.  However, the significance is greater than 3$\sigma$ when
 using the RMS scatter of a blank portion of the composite spectrum as the error spectrum.  Past studies have used this
 technique to estimate uncertainties on emission-line fluxes from stacked spectra \citep{erb06,and13}.  Given our very conservative process for
 estimating uncertainties, and in order to be comparable to other works, we treat all measurements of the composite spectra
 that have significance greater than 3$\sigma$ when using the RMS scatter as detections, and as limits otherwise.
  This practice is adopted throughout this paper.  Plotted error bars still denote the uncertainty estimated by the process
 described above.

\begin{table*}[t]
 \caption{Galaxy properties and emission-line luminosites from $z\sim2.3$ composite spectra
 }\label{tab:binvalues}
 \begin{tabular*}{\textwidth}{@{\extracolsep{\fill}} l l c c c c c c c c }
   \hline\hline
   $\log{(\frac{\mbox{M}_*}{\mbox{M}_\odot})}$\tablenotemark{a} & $\langle \log{(\frac{\mbox{M}_*}{\mbox{M}_\odot})}\rangle$\tablenotemark{b} & N$_{\mbox{gal}}$\tablenotemark{c} & SFR$_{\mbox{med}}$\tablenotemark{d} & L$_{\mbox{[N~\textsc{ii}]}}$\tablenotemark{e} & L$_{\mbox{H}\alpha}$\tablenotemark{e} & L$_{\mbox{[O~\textsc{iii}]}}$\tablenotemark{e} & L$_{\mbox{H}\beta}$\tablenotemark{e} & \multicolumn{2}{c}{$12+\log{(\mbox{O/H})}$} \\
   \cline{9-10}
     &   &  & {\scriptsize ($\msun\mbox{ yr}^{-1}$)} & {\scriptsize ($10^{42}$ erg s$^{-1}$)} & {\scriptsize ($10^{42}$ erg s$^{-1}$)} & {\scriptsize ($10^{42}$ erg s$^{-1}$)} & {\scriptsize ($10^{42}$ erg s$^{-1}$)} & N2\tablenotemark{f} & O3N2\tablenotemark{g} \\
   \hline
   \hline
   \multicolumn{10}{c}{Full sample} \\
   \hline
   9.15-9.68 & 9.45 & 22 & 11.6 & $0.11\pm0.04$ & $2.03\pm0.05$ & $2.78\pm0.14$ & $0.60\pm0.03$ & $8.18^{+0.07}_{-0.10}$ & $8.11^{+0.04}_{-0.06}$ \\
   9.68-9.94 & 9.84 & 22 & 23.4 & $0.27\pm0.05$ & $3.03\pm0.08$ & $3.21\pm0.14$ & $0.82\pm0.05$ & $8.30^{+0.04}_{-0.05}$ & $8.20^{+0.02}_{-0.03}$ \\
   9.99-10.27 & 10.11 & 22 & 26.8 & $0.49\pm0.08$ & $3.05\pm0.10$ & $2.44\pm0.20$ & $0.76\pm0.06$ & $8.44^{+0.04}_{-0.04}$ & $8.31^{+0.02}_{-0.03}$ \\
   10.29-11.11 & 10.56 & 21 & 53.8 & $1.07\pm0.10$ & $4.82\pm0.12$ & $2.01\pm0.21$ & $0.98\pm0.08$ & $8.52^{+0.02}_{-0.02}$ & $8.42^{+0.02}_{-0.02}$ \\
   \hline 
  \multicolumn{10}{c}{High-SFR subsample} \\
   \hline
   9.23-9.89 & 9.70 & 11 & 37.1 & $0.26\pm0.08$ & $3.61\pm0.05$ & $4.35\pm0.20$ & $0.83\pm0.08$ & $8.25^{+0.06}_{-0.08}$ & $8.13^{+0.03}_{-0.05}$ \\
   9.90-10.11 & 10.02 & 11 & 30.5 & $0.58\pm0.11$ & $3.49\pm0.08$ & $2.72\pm0.25$ & $0.79\pm0.09$ & $8.45^{+0.04}_{-0.05}$ & $8.31^{+0.03}_{-0.03}$ \\
   10.13-10.45 & 10.33 & 11 & 74.2 & $0.80\pm0.19$ & $5.46\pm0.10$ & $2.92\pm0.30$ & $1.09\pm0.14$ & $8.42^{+0.05}_{-0.06}$ & $8.32^{+0.03}_{-0.04}$ \\
   10.46-11.11 & 10.73 & 11 & 41.2 & $1.09\pm0.12$ & $4.71\pm0.12$ & $1.88\pm0.27$ & $0.89\pm0.11$ & $8.53^{+0.02}_{-0.02}$ & $8.42^{+0.03}_{-0.03}$ \\
   \hline
  \multicolumn{10}{c}{Low-SFR subsample} \\
   \hline
   9.15-9.47 & 9.32 & 11 & 11.1 & $0.09\pm0.04$ & $1.83\pm0.08$ & $2.69\pm0.22$ & $0.57\pm0.05$ & $8.28$\tablenotemark{h} & $8.16$\tablenotemark{h} \\
   9.52-9.75 & 9.65 & 11 & 9.26 & $0.14\pm0.08$ & $1.82\pm0.07$ & $2.40\pm0.10$ & $0.50\pm0.05$ & $8.27^{+0.12}_{-0.16}$ & $8.16^{+0.07}_{-0.09}$ \\
   9.80-10.02 & 9.89 & 11 & 14.5 & $0.22\pm0.08$ & $2.59\pm0.09$ & $2.68\pm0.30$ & $0.76\pm0.07$ & $8.29^{+0.07}_{-0.10}$ & $8.21^{+0.04}_{-0.06}$ \\
   10.07-10.40 & 10.21 & 10 & 11.7 & $0.43\pm0.09$ & $2.19\pm0.10$ & $1.44\pm0.13$ & $0.62\pm0.07$ & $8.50^{+0.05}_{-0.06}$ & $8.38^{+0.03}_{-0.04}$ \\
   \hline \\
 \end{tabular*}
 \tablenotetext{1}{Range of $\log{(\mbox{M}_*/\msun)}$ of galaxies in a bin.}
 \tablenotetext{2}{Average $\log{(\mbox{M}_*/\msun)}$ of galaxies in a bin.}
 \tablenotetext{3}{Number of galaxies in a bin.}
 \tablenotetext{4}{Median dust-corrected H$\alpha$ SFR of galaxies in a bin.}
 \tablenotetext{5}{Emission-line luminosity and uncertainty on [N~\textsc{ii}]$\lambda$6584, H$\alpha$, [O~\textsc{iii}]$\lambda$5007, and H$\beta$,
 as measured from the composite spectra.}
 \tablenotetext{6}{Oxygen abundance and uncertainty determined with the N2 indicator using equation~(\ref{eq:pp04n2}).}
 \tablenotetext{7}{Oxygen abundance and uncertainty determined with the O3N2 indicator using equation~(\ref{eq:pp04o3n2}).}
 \tablenotetext{8}{3$\sigma$ upper limit on the oxygen abundance where [N~\textsc{ii}]$\lambda$6584 is not detected.}
\end{table*}

Measurements from the four mass bins are shown in green in Figure~\ref{fig:mzr}.  Bin points are plotted
 at the average $\log($\mstar$/\msun)$, the vertical error bar is the uncertainty in the oxygen abundance, and the horizontal
 error bar shows the range of stellar masses in that bin.  The calibration uncertainty is not included in the
 metallicity uncertainty.  Note that the calibration uncertainty should be reduced by a factor of $\sqrt{\mbox{N}}$ when
 using stacked spectra, where N is the number of galaxies in the stack \citep{erb06}.  The reduction in the calibration
 uncertainty is approximately a factor of $\sqrt{22}\approx4.7$ for our mass bins, yielding binned calibration
 uncertainties of 0.038 and 0.030 dex respectively for the N2 and O3N2 calibrations.

After binning $z\sim2.3$ MOSDEF galaxies according to stellar mass, we find a clear progression in which metallicity
 increases monotonically as stellar mass increases, in agreement with previous studies \citep{erb06,mai08,ste14}.
  This progression is well described by a single power law when determining metallicities
 with the O3N2 indicator, while it appears to flatten at high stellar masses when metallicities are based on
 the N2 indicator.  We note that the two indicators yield different values for the low-mass slope of the MZR,
 which is an important test of outflow models.  This difference is further evidence that care must be taken interpreting results
 that are dependent on the metallicity calibration used \citep{kew08,and13}, and demonstrates that disagreement between
 metallicity indicators persists at high redshifts.  We find that the $z\sim2.3$ MZR is offset below the local MZR
 by $\sim0.15-0.3$ dex and $\sim0.3$ dex based on the N2 and O3N2 indicators, respectively.  The offset observed
 with the N2 indicator is very similar to that found by \citet{erb06} using stacked spectra of $z\sim2.2$ galaxies
 and the \citet{pet04} N2 metallicity calibration.
  However, the N2 indicator must be used with caution at high redshifts due to secondary dependences
 on the ionization parameter, N/O abundance ratio, and hardness of the ionizing spectrum, some or all of which may
 evolve with redshift.
  In addition to these parameters, the N2 line ratio can be significantly affected by the presence of
 shock excitation which could be present in high-redshift galaxies due to large gas flows \citep{new14}.
  It is likely that the true offset in the N2 MZR is larger than that shown in Figure~\ref{fig:mzr}
 since the N2 indicator is believed to overestimate the metallicity at high redshifts \citep{liu08,new14}.
  While changes in parameters such as the N/O abundance ratio could also bias metallicity estimates
 of the O3N2 indicator, \citet{liu08} and \citet{ste14} have found that O3N2 is significantly less biased 
 than the N2 indicator.
  One aspect of the $z\sim2.3$ MOSDEF sample is the requirement of both H$\alpha$ and H$\beta$ detections
 in order to estimate dust-corrected SFRs.
  We tested that the S/N requirement for H$\beta$ does not bias the $z\sim2.3$ sample
 against dusty metal-rich galaxies by including
 galaxies with H$\beta$ upper limits in the composite spectra.  Emission line measurements from such composite spectra
 agreed with those presented in Table~\ref{tab:binvalues} to better than 1$\sigma$ and displayed no systematic offset.

\begin{figure}
 \centering
 \includegraphics[width=\columnwidth]{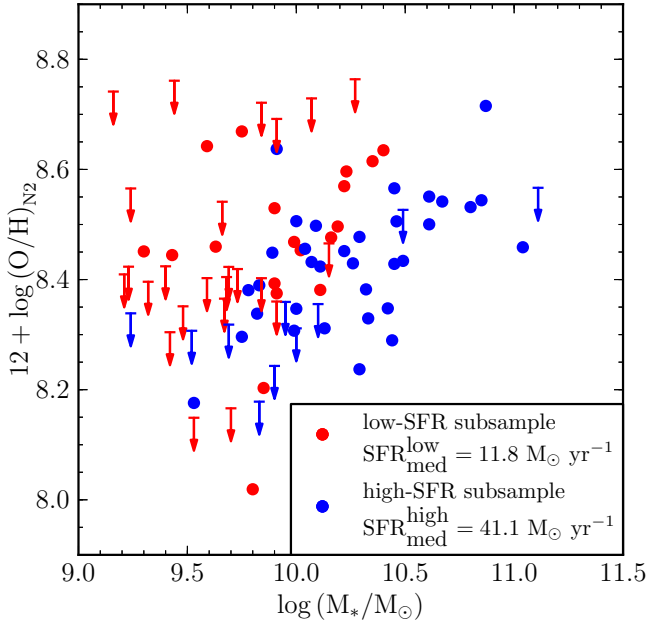}
 \caption{SFR dependence in the MZR at $z\sim2.3$.
  Individual galaxies in the $z\sim2.3$ sample are divided into high- and low-SFR subsamples at the median SFR
 of $25.9\ \msun\mbox{ yr}^{-1}$.  The high- and low-SFR subsamples have median SFRs of
 11.8~$\msun\mbox{ yr}^{-1}$ and 41.1~$\msun\mbox{ yr}^{-1}$, respectively.
}\label{fig:sfrcolors}
\end{figure}

In order to investigate the SFR dependence of the $z\sim2.3$ MZR, we divide the sample into high-SFR and low-SFR
 subsamples at the median SFR, as shown in Figure~\ref{fig:sfrcolors}.  The median SFR of the total
 sample is 25.9~$\msun\mbox{ yr}^{-1}$, while the median SFRs of the high- and low-SFR subsamples are
 41.1~$\msun\mbox{ yr}^{-1}$ and 11.8~$\msun\mbox{ yr}^{-1}$, respectively.
  The abundance of upper limits, especially
 in the low-\mstar, low-SFR regime, makes it difficult to determine if SFR dependence is present.
  There is only a narrow mass range of $\log($\mstar$/\msun)\sim10.0-10.5$ that is populated by detections
 from both the high- and low-SFR
 subsample without a significant number of limits.  Although this region may appear
 to suggest SFR dependence among individual
 $z\sim2.3$ galaxies, a larger dynamic range in stellar mass is needed to confirm any trend.
  Additionally, much of the division seen between the high- and low-SFR subsamples in Figure~\ref{fig:sfrcolors} is
 a manifestation of the \mstar-SFR relation for star-forming galaxies, according to which lower SFRs are more common
 among low-mass galaxies.
  Indeed, dividing a sample by SFR alone results in an offset between
the average stellar masses for the two SFR subsamples (see Figure~\ref{fig:sfrcolors}),
with high-SFR objects characterized by higher stellar masses on average
than those in the low-SFR subsample.
  In order to overcome the difficulty of interpreting upper limits in $12+\log(\mbox{O/H})$
 and avoid stellar mass selection effects,
 we created composite spectra
 in four bins of stellar mass for each SFR subsample.
  Binning in both SFR and stellar mass is equivalent to selecting galaxies with a narrow range of
 sSFR, a property which has a weaker mass dependence than SFR.
  The median SFRs
 of the high-SFR bins range from $37.1-74.2~\msun\mbox{ yr}^{-1}$, while the median SFRs of the low-SFR bins range from
 $9.3-14.5~\msun\mbox{ yr}^{-1}$.
  Composite spectra were produced by applying the same binning and stacking process
 outlined above to each subsample.  Emission-line measurements and metallicity estimates were obtained in the same
 manner as before.  Bin properties, emission-line measurements, and oxygen abundances for the high- and low-SFR
 subsamples are presented in Table~\ref{tab:binvalues}.

\begin{figure*}[t]
        \centering
        \includegraphics[width=\textwidth]{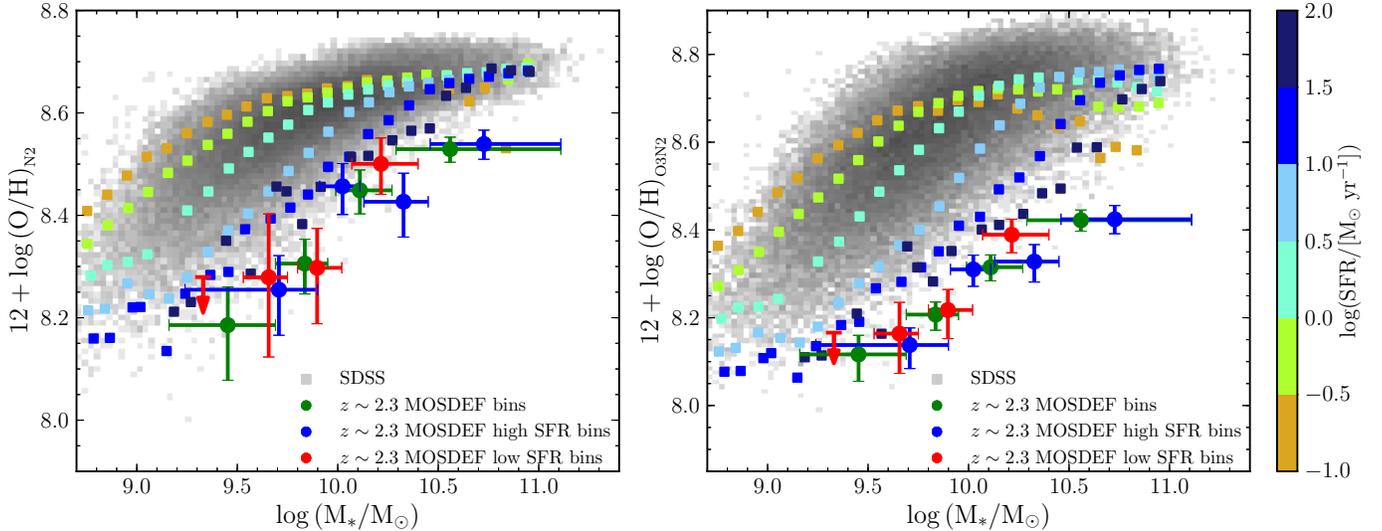}
        \caption{Comparison of \mstar, metallicity, and SFR between $z\sim2.3$ and local galaxies.
  The $z\sim2.3$ high- and low-SFR subsamples are separated into four stellar
 mass bins, with metallicities determined using the N2 (left) and O3N2 (right) indicators.
  The blue and red points and error bars indicate bins of the high- and low-SFR subsamples, respectively.  The green
 points with error bars indicate stellar mass bins from the full sample.  Error bars for all binned points are the same
 as in Figure~\ref{fig:mzr}.  The gray
 two-dimensional histogram shows the density of local SDSS galaxies in this parameter space.  Colored squares
 are \mstar-SFR bins of local SDSS star-forming galaxies from \citet{and13}, with the color indicating the range of
 SFRs in a bin (see colorbar).  Red MOSDEF $z\sim2.3$ bins are comparable
 to SDSS bins with $\log{(\mbox{SFR})}=1.0-1.5$ (medium blue), while blue MOSDEF bins are comparable to those
 with $\log{(\mbox{SFR})}=1.5-2.0$ (dark blue).
	}\label{fig:mzrandmart}
\end{figure*}

In Figure~\ref{fig:mzrandmart}, the high-SFR, low-SFR, and full sample bins are shown in blue, red, and green,
 respectively, with error bars as in Figure~\ref{fig:mzr}.  Horizontal bars with downward arrows denote 3$\sigma$
 upper limits where [N~\textsc{ii}]$\lambda$6584 was not detected.
  For a comparison to the local universe, we use measurements from the stacked SDSS spectra of
 \citet{and13}.  These stacks constitute a fair comparison to the $z\sim2.3$ MOSDEF stacks as the galaxies are
 also binned according to \mstar~and SFR and the spectra are combined in a very similar manner to our method.
  We use published emission-line measurements for the \citet{and13} stacks to estimate the metallicities
 using equations~(\ref{eq:pp04n2}) and~(\ref{eq:pp04o3n2}).  The SDSS stacks are shown in Figure~\ref{fig:mzrandmart}
 as squares, where the color represents the SFR range of that bin.  The SDSS stacks can be compared directly to
 the $z\sim2.3$ stacks as they are produced using the same methods, same metallicity calibrations,
 and consistent SFR estimates.

We do not see evidence of significant SFR dependence in the $z\sim2.3$ MZR.
  Bins of the high- and low-SFR subsamples do not follow a different MZR from that of the full sample bins
 within the uncertainties.  Error bars of the high- and low-SFR subsample bins overlap with the full sample and
 each other, and the SFR subsamples appear to scatter about the full sample.
While there may still be SFR dependence of the MZR at $z\sim2.3$, our current sample
 lacks the size and possibly the dynamic range required to resolve it.  By comparing the \citet{and13} SDSS
 \mstar-SFR bins to the $z\sim2.3$ MOSDEF bins, we find that $z\sim2.3$ galaxies have lower metallicities at a given
 \mstar~and SFR than is observed locally, with the high-redshift bins being offset $\sim$0.1~dex below the local bins.
  This result is confirmed using both the N2 and O3N2 indicators.
  According to the median SFRs of the MOSDEF bins, the low-SFR subsample matches
 the medium blue SDSS bins ($10-31.6~\msun\mbox{ yr}^{-1}$) and the high-SFR subsample matches the dark blue SDSS bins
 ($31.6-100~\msun\mbox{ yr}^{-1}$).
  Due to the increase with redshift of the typical sSFR at a given stellar mass, we can only compare
 the $z\sim2.3$ MOSDEF sample with the high-SFR tail of the local distribution of star-forming galaxies.
  However, given the large size of the local sample, this high-SFR tail contains a sufficient number galaxies
 for a robust comparison ($\sim10^4$).

  Given that there are SDSS bins across the entire range of stellar masses probed
 by the MOSDEF sample, the MOSDEF bins should have the same metallicities as the SDSS bins with comparable \mstar~and SFR
 if the local FMR holds at this redshift.  This is not the case.  We conclude that $z\sim2.3$ star-forming galaxies do not
 lie on the local FMR.  An alternate explanation of this offset is that local metallicity calibrations do not hold at high
 redshifts, discussed further in Section~\ref{sec:disc}.  However, there is evidence that the N2 indicator may
 overestimate the oxygen abundance in high-redshift galaxies \citep{liu08,new14}, in which case the true offset would be larger
 than the one displayed in Figure~\ref{fig:mzrandmart}, strengthening the claim that $z\sim2.3$ galaxies do not
 fall on the local FMR.  The O3N2 indicator is not expected to be significantly affected by redshift evolution \citep{ste14}.

\section{Summary and Discussion}\label{sec:disc}

In this paper, we used early observations from the MOSDEF survey to investigate the $z\sim2.3$ mass-metallicity relationship.  Results were based on 87 galaxies with individual measurements from a rest-frame optical selected
 sample with coverage of all strong optical emission lines.  We find a clear positive correlation between
 \mstar~and metallicity using composite spectra of galaxies binned by stellar mass.  At this point, we defer measurements of
 the scatter and slope of the $z\sim2.3$ MZR due to uncertainty regarding the reliability of local metallicity calibrations at
 high redshifts, discussed below.

We investigated the SFR dependence of the $z\sim2.3$ MZR by dividing the sample
 at the median SFR and making composite spectra of galaxies binned according to stellar mass within the high- and low-SFR
 subsamples.  We do not observe a significant dependence of metallicity on SFR at a given \mstar.  However, there is not
 strong SFR dependence of metallicity within local SDSS galaxies at comparable SFRs, as seen in the
 medium and dark blue squares in Figure~\ref{fig:mzrandmart}.
  Given the uncertainties in oxygen abundance measurements for the $z\sim2.3$ SFR bins, we are unable
 to resolve SFR dependence at the level that is observed in bins of local galaxies.
  Larger samples at $z\sim2.3$ will be required to confirm or rule out such SFR dependence.

An outstanding question
 in galaxy evolution is whether or not high redshift galaxies fall on the local FMR.
  If the FMR is universal and redshift independent, then high- and low-redshift galaxies have similar metallicity equilibrium
 conditions for the balance between gas inflows and outflows, and star formation.
  To address this question, we compared
 the $z\sim2.3$ MOSDEF stacks to composite spectra of local star-forming galaxies from \citet{and13}.
  We find $z\sim2.3$ star-forming galaxies are $\sim0.1$ dex lower in
 metallicity for a given \mstar~and SFR than the local FMR predicts, in agreement with some high redshift
 studies \citep{zah13,cul14,tro14,wuy14}.
  Other studies have found agreement with the local FMR at these redshifts \citep[e.g.,][]{bel13}.

Of key importance to our study is the ability to directly compare SFR, stellar mass, and metallicity between
 the $z\sim2.3$ MOSDEF sample and the local comparison sample.
  The SFRs used by \citet{and13} were estimated following \citet{bri04} which utilizes multiple
 emission lines simultaneously to estimate the SFR, but heavily weights H$\alpha$ and H$\beta$, and is thus
 consistent with SFRs estimated using dust-corrected H$\alpha$.  We have independently confirmed this consistency
 with SDSS DR7 measurements.  Furthermore, both MOSDEF and SDSS SFRs are corrected to total galaxy SFRs,
 with estimates for slit losses in the case of MOSDEF, and fiber losses in the case of SDSS.
  Stellar masses for both MOSDEF and SDSS galaxies\footnote[8]{Stellar mass estimates from the
 MPA-JHU SDSS DR7 spectroscopic catalog are based on fits to the photometry rather than spectral indices of stellar
 absorption features which were used for previous releases.  See \texttt{http://www.mpa-garching.mpg.de/SDSS/DR7/mass\_comp.html}
 for a comparison of SDSS stellar masses based on indices and photometry.} are based on SED-fitting to broadband photometry.
  Finally, we use a stacking procedure nearly identical to that of \citet{and13} and estimate metallicity using the same
 indicators and calibrations for each dataset.  In summary, our comparison to the
 \citet{and13} stacks constitutes a fair and direct FMR comparison because the two samples use the same
 metallicity calibrations and methods for stacking galaxy spectra, as well as consistent SFR and stellar mass estimates.

  One difference between the $z\sim2.3$ MODSEF sample and the local comparison sample is the method of
 obtaining spectra.  MOSDEF data are obtained by placing a $0\secpoint7$ slit on the target which typically contains
 a large fraction of the total light from the galaxy, while SDSS spectra are obtained by placing a 3"-diameter
 fiber on the centers of galaxies.  Measured metallicities can be sensitive to the method of obtaining spectra
 if radial metallicity gradients are present.  In the local universe, star-forming galaxies exhibit negative radial metallicity
 gradients such that the inner regions of galaxies (probed by SDSS fibers) are more
 metal-rich than the outer regions \citep[e.g.,][]{vil92}.
  At $z>2$, observations have not yet confirmed the existence
 of ubiquitous metallicity gradients among star-forming galaxies, with various groups reporting negative, flat, or
 even positive (inverted) metallicity gradients \citep{cre10,jon10,jon13,que12,sto14}.
  It is not currently possible to estimate how
 metallicity measurements at high redshifts may be biased because of metallicity gradients.  However, due to the
 optimal extraction method we used, the line
 ratios measured for the MOSDEF sample are dominated by light from the inner regions of galaxies where the surface brightness
 is greatest.  Thus, both MOSDEF and SDSS measure the metallicity of the innermost regions of star-forming galaxies.

An additional strength of our comparison is that it does not depend on any extrapolation
 of a parameterization of the local FMR and is thus free
 of the effects that the choice of extrapolation used can have on conclusions regarding the FMR, described in \citet{mai14}.
  We emphasize that proper investigation of the universality of the FMR with redshift
 requires \textit{both} checking for
 consistency with the local FMR or its projections \textit{and} looking for SFR dependence within the high-redshift sample.
  Many previous studies have overlooked SFR dependence within the sample, or have been unable to investigate this
 aspect of the FMR
 due to small or incomplete samples.  We have done both in this paper, and additionally used dust-corrected H$\alpha$ SFRs
 which are independent of the SED fitting used to determine stellar masses.
  A consistency of the bulk properties of a high-redshift
 sample with the local FMR is not sufficient proof that the relationship between SFR, \mstar, and metallicity is the
 same at high redshifts.

If the observed $z\sim2.3$ offset from the local FMR is real and not an artifact arising from unreliable
 metallicity calibrations at high redshifts, it may be evidence of the ``gas accumulation phase" described by
 \citet{dav12}.  This phase occurs during galaxy growth when a galaxy cannot process inflowing gas and form stars
 as quickly as gas is accreted, building up the gas reservoir.  In this case, metallicities are lower than expected
 at a given \mstar~and SFR because the ISM metallicity is diluted faster than metals are produced in stars.
  Large accretion rates can cause this imbalance,
 suggesting the possibility that the environments of $z\sim2.3$ star-forming galaxies lead to high gas accretion rates.
  There is some evidence in the literature of extreme accretion rates  at $z\gtrsim2$, as suggested by gas mass fractions
 and sSFR \citep{tac10,tac13,red12}.
  However, the end of the gas accumulation phase is predicted to occur at $z\gtrsim4$ \citep{dav12},
 as has been suggested by models of the star-formation histories of Lyman-break galaxies \citep{pap11}.

\begin{figure}
 \centering
 \includegraphics[width=\columnwidth]{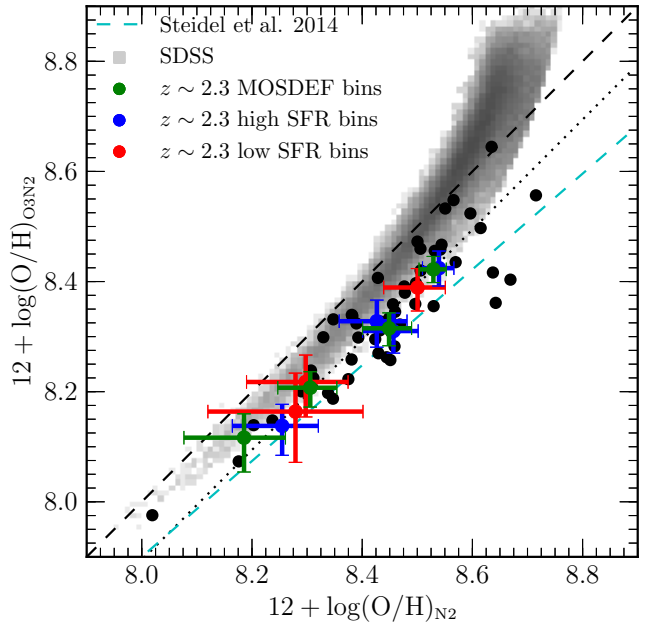}
 \caption{Comparison of metallicities estimated using the N2 and O3N2 indicators.  The gray blocks show the density of
 local SDSS galaxies.  Individual $z\sim2.3$ galaxies with detections in H$\beta$, [O~\textsc{iii}]$\lambda$5007, H$\alpha$, and
 [N~\textsc{ii}]$\lambda$6584 are indicated by black points.  Green points and error bars represent stellar mass bins of the full
 $z\sim2.3$ sample, while blue and red points and error bars indicate the high- and low-SFR subsamples, respectively.  The
 black dashed line indicates a one-to-one correspondence.  The dotted line is the best-fit line of slope unity through the individual
 MOSDEF galaxies, offset 0.1 dex below the one-to-one line.
  The cyan dashed line indicates the best-fit line to $z\sim2.3$ star-forming galaxies from \citet{ste14}.
}\label{fig:metcomp}
\end{figure}

We present these results with one very important caveat.  Accurately determining metallicities at different redshifts
 is of key importance to studying the evolution of the MZR.  In the local universe, relationships between strong
 emission line ratios and metallicity can be calibrated to ``direct" electron temperature-determined metallicities
 from measuring auroral lines such as [O~\textsc{iii}]$\lambda$4363 \citep{pet04,pil05} or photoionization models of
 star-forming regions \citep{zar94,kew02,kob04,tre04}.  At redshifts above $z\sim1$, it is nearly impossible to detect weak
 auroral lines for directly determining metallicity \citep[but see][]{yua09,rig11,bra12b,chr12,mas14}.
  Creating photoionization models that suitably represent high-redshift star-forming regions requires knowledge of
 physical parameters which have been poorly constrained up to this point.  Thus, it is unknown if local metallicity
 calibrations hold at high redshifts.  Figure~\ref{fig:metcomp} shows a comparison between metallicities
 determined using the O3N2 indicator and the N2 indicator for both local SDSS galaxies (grey points) and MOSDEF $z\sim2.3$
 galaxies (black points).  The black dashed line indicates a one-to-one relationship.  If local calibrations do indeed hold
 at high redshifts, then the relationship between metallicities determined from different indicators should not evolve with
 redshift.  It is clear that the $z\sim2.3$ galaxies
 are offset below the local galaxies.  The dotted line is the best-fit line of slope unity to the individual $z\sim2.3$ galaxies,
 yielding an offset of -0.1 dex from a one-to-one correspondence, over twice that displayed by the SDSS sample.
  \citet{ste14} found an offset slightly larger than this at $z\sim2.3$.
  This offset
 demonstrates that the two metallicity indicators are not evolving in the same way with redshift, and shows the
 need of metallicity calibrations appropriate for high redshift galaxies.

  There is mounting evidence in the literature that high-redshift star-forming galaxies have markedly different
 emission line ratios from those of local star-forming galaxies \citep{sha05,liu08,hai09,mast14,ste14}.  This has been observed as an offset of the
 star-forming sequence in the BPT diagram for high-redshift galaxies \citep{sha14}.
  The difference in diagnostic emission
 line ratios suggests that the physical conditions of high-redshift star-forming regions are different from what
 is seen locally.  If true, photoionization models of local H~\textsc{ii} regions are unsuitable to describe their high-redshift
 counterparts and one would expect the relationship between metallicity and strong emission line ratios to differ.
  One avenue forward is to use emission line ratios to constrain the physical conditions of high-redshift
 star-forming regions with a statistically significant sample and re-calibrate to photoionization models using these
 new constraints as input parameters.  When complete, the MOSDEF survey will provide a sample of galaxies in three redshift
 bins spanning $z\sim1.5-3.5$ with rest-frame optical spectra covering all strong optical emission lines that is an order of
 magnitude larger than similar existing samples.  Using this dataset, we will constrain the physical conditions of high-redshift
 star-forming regions in order to provide input parameters for photoionization models that are appropriate for these redshifts.
  Predictions from these models can then be used to produce new metallicity calibrations that are suitable for high redshift
 galaxies that have more extreme interstellar media and star-forming regions than typically observed in the local universe.
  We will also consider additional line ratios used as metallicity indicators in order to more fully understand the
 bias of local metallicity calibrations at high redshifts, a critical step to estimating reliable abundances at
 these redshifts.
  Such robust metallicities are required to accurately measure the evolution, slope, and scatter of the MZR and
 investigate the existence of the FMR at high redshifts and, by extension, uncover the nature of gas flows at high redshifts.

\acknowledgements We acknowledge support from NSF AAG grants AST-1312780, 1312547, 1312764, and 1313171, and grant AR-13907
from the Space Telescope Science Institute.
 Support for program AR-13907 was provided by NASA through a grant from the Space Telescope Science
 Institute, which is operated by the Association of Universities for Research in Astronomy, Inc.,
 under NASA contract NAS 5-26555.
 We are also grateful to Marc Kassis at the Keck Observatory for his many valuable contributions to
the execution of the MOSDEF survey. We also acknowledge
the 3D-HST collaboration, who provided us with spectroscopic
and photometric catalogs used to select MOSDEF targets and derive
stellar population parameters. We also thank I. McLean, K. Kulas,
and G. Mace for taking observations for the MOSDEF survey in May and June 2013.
MK acknowledges support from a Committee Faculty Research Grant and a Hellmann Fellowship. ALC
acknowledges funding from NSF CAREER grant AST-1055081. NAR is supported by
an Alfred P. Sloan Research Fellowship.
We wish to extend special thanks to those of Hawaiian ancestry on
whose sacred mountain we are privileged to be guests. Without their generous hospitality, most
of the observations presented herein would not have been possible.

\bibliography{mzrpaper}

\begin{thebibliography}{}
\expandafter\ifx\csname natexlab\endcsname\relax\def\natexlab#1{#1}\fi

\bibitem[{{Abazajian} {et~al.}(2009){Abazajian}, {Adelman-McCarthy},
  {Ag{\"u}eros}, {Allam}, {Allende Prieto}, {An}, {Anderson}, {Anderson},
  {Annis}, {Bahcall}, \& et~al.}]{aba09}
{Abazajian}, K.~N., {Adelman-McCarthy}, J.~K., {Ag{\"u}eros}, M.~A., {et~al.}
  2009, \apjs, 182, 543

\bibitem[{{Andrews} \& {Martini}(2013)}]{and13}
{Andrews}, B.~H., \& {Martini}, P. 2013, \apj, 765, 140

\bibitem[{{Baldwin} {et~al.}(1981){Baldwin}, {Phillips}, \&
  {Terlevich}}]{bpt81}
{Baldwin}, J.~A., {Phillips}, M.~M., \& {Terlevich}, R. 1981, \pasp, 93, 5

\bibitem[{{Belli} {et~al.}(2013){Belli}, {Jones}, {Ellis}, \&
  {Richard}}]{bel13}
{Belli}, S., {Jones}, T., {Ellis}, R.~S., \& {Richard}, J. 2013, \apj, 772, 141

\bibitem[{{Brammer} {et~al.}(2012{\natexlab{a}}){Brammer}, {van Dokkum},
  {Franx}, {Fumagalli}, {Patel}, {Rix}, {Skelton}, {Kriek}, {Nelson},
  {Schmidt}, {Bezanson}, {da Cunha}, {Erb}, {Fan}, {F{\"o}rster Schreiber},
  {Illingworth}, {Labb{\'e}}, {Leja}, {Lundgren}, {Magee}, {Marchesini},
  {McCarthy}, {Momcheva}, {Muzzin}, {Quadri}, {Steidel}, {Tal}, {Wake},
  {Whitaker}, \& {Williams}}]{bra12a}
{Brammer}, G.~B., {van Dokkum}, P.~G., {Franx}, M., {et~al.}
  2012{\natexlab{a}}, \apjs, 200, 13

\bibitem[{{Brammer} {et~al.}(2012{\natexlab{b}}){Brammer},
  {S{\'a}nchez-Janssen}, {Labb{\'e}}, {da Cunha}, {Erb}, {Franx}, {Fumagalli},
  {Lundgren}, {Marchesini}, {Momcheva}, {Nelson}, {Patel}, {Quadri}, {Rix},
  {Skelton}, {Schmidt}, {van der Wel}, {van Dokkum}, {Wake}, \&
  {Whitaker}}]{bra12b}
{Brammer}, G.~B., {S{\'a}nchez-Janssen}, R., {Labb{\'e}}, I., {et~al.}
  2012{\natexlab{b}}, \apjl, 758, L17

\bibitem[{{Brinchmann} {et~al.}(2004){Brinchmann}, {Charlot}, {White},
  {Tremonti}, {Kauffmann}, {Heckman}, \& {Brinkmann}}]{bri04}
{Brinchmann}, J., {Charlot}, S., {White}, S.~D.~M., {et~al.} 2004, \mnras, 351,
  1151

\bibitem[{{Calzetti} {et~al.}(2000){Calzetti}, {Armus}, {Bohlin}, {Kinney},
  {Koornneef}, \& {Storchi-Bergmann}}]{cal00}
{Calzetti}, D., {Armus}, L., {Bohlin}, R.~C., {et~al.} 2000, \apj, 533, 682

\bibitem[{{Chabrier}(2003)}]{cha03}
{Chabrier}, G. 2003, \pasp, 115, 763

\bibitem[{{Christensen} {et~al.}(2012){Christensen}, {Laursen}, {Richard},
  {Hjorth}, {Milvang-Jensen}, {Dessauges-Zavadsky}, {Limousin}, {Grillo}, \&
  {Ebeling}}]{chr12}
{Christensen}, L., {Laursen}, P., {Richard}, J., {et~al.} 2012, \mnras, 427,
  1973

\bibitem[{{Coil} {et~al.}(2014){Coil}, {Aird}, {Reddy}, {Shapley}, {Kriek},
  {Siana}, {Mobasher}, {Freeman}, {Price}, \& {Shivaei}}]{coi14}
{Coil}, A.~L., {Aird}, J., {Reddy}, N., {et~al.} 2014, ArXiv e-prints,
  arXiv:1409.6522

\bibitem[{{Conroy} {et~al.}(2009){Conroy}, {Gunn}, \& {White}}]{con09}
{Conroy}, C., {Gunn}, J.~E., \& {White}, M. 2009, \apj, 699, 486

\bibitem[{{Cresci} {et~al.}(2010){Cresci}, {Mannucci}, {Maiolino}, {Marconi},
  {Gnerucci}, \& {Magrini}}]{cre10}
{Cresci}, G., {Mannucci}, F., {Maiolino}, R., {et~al.} 2010, \nat, 467, 811

\bibitem[{{Cullen} {et~al.}(2014){Cullen}, {Cirasuolo}, {McLure}, {Dunlop}, \&
  {Bowler}}]{cul14}
{Cullen}, F., {Cirasuolo}, M., {McLure}, R.~J., {Dunlop}, J.~S., \& {Bowler},
  R.~A.~A. 2014, \mnras, 440, 2300

\bibitem[{{Dav{\'e}} {et~al.}(2011){Dav{\'e}}, {Finlator}, \&
  {Oppenheimer}}]{dav11}
{Dav{\'e}}, R., {Finlator}, K., \& {Oppenheimer}, B.~D. 2011, \mnras, 416, 1354

\bibitem[{{Dav{\'e}} {et~al.}(2012){Dav{\'e}}, {Finlator}, \&
  {Oppenheimer}}]{dav12}
---. 2012, \mnras, 421, 98

\bibitem[{{Dekel} \& {Silk}(1986)}]{dek86}
{Dekel}, A., \& {Silk}, J. 1986, \apj, 303, 39

\bibitem[{{Donley} {et~al.}(2012){Donley}, {Koekemoer}, {Brusa}, {Capak},
  {Cardamone}, {Civano}, {Ilbert}, {Impey}, {Kartaltepe}, {Miyaji}, {Salvato},
  {Sanders}, {Trump}, \& {Zamorani}}]{don12}
{Donley}, J.~L., {Koekemoer}, A.~M., {Brusa}, M., {et~al.} 2012, \apj, 748, 142

\bibitem[{{Erb} {et~al.}(2006){Erb}, {Shapley}, {Pettini}, {Steidel}, {Reddy},
  \& {Adelberger}}]{erb06}
{Erb}, D.~K., {Shapley}, A.~E., {Pettini}, M., {et~al.} 2006, \apj, 644, 813

\bibitem[{{Finlator} \& {Dav{\'e}}(2008)}]{fin08}
{Finlator}, K., \& {Dav{\'e}}, R. 2008, \mnras, 385, 2181

\bibitem[{{Grogin} {et~al.}(2011){Grogin}, {Kocevski}, {Faber}, {Ferguson},
  {Koekemoer}, {Riess}, {Acquaviva}, {Alexander}, {Almaini}, {Ashby}, {Barden},
  {Bell}, {Bournaud}, {Brown}, {Caputi}, {Casertano}, {Cassata}, {Castellano},
  {Challis}, {Chary}, {Cheung}, {Cirasuolo}, {Conselice}, {Roshan Cooray},
  {Croton}, {Daddi}, {Dahlen}, {Dav{\'e}}, {de Mello}, {Dekel}, {Dickinson},
  {Dolch}, {Donley}, {Dunlop}, {Dutton}, {Elbaz}, {Fazio}, {Filippenko},
  {Finkelstein}, {Fontana}, {Gardner}, {Garnavich}, {Gawiser}, {Giavalisco},
  {Grazian}, {Guo}, {Hathi}, {H{\"a}ussler}, {Hopkins}, {Huang}, {Huang},
  {Jha}, {Kartaltepe}, {Kirshner}, {Koo}, {Lai}, {Lee}, {Li}, {Lotz}, {Lucas},
  {Madau}, {McCarthy}, {McGrath}, {McIntosh}, {McLure}, {Mobasher},
  {Moustakas}, {Mozena}, {Nandra}, {Newman}, {Niemi}, {Noeske}, {Papovich},
  {Pentericci}, {Pope}, {Primack}, {Rajan}, {Ravindranath}, {Reddy}, {Renzini},
  {Rix}, {Robaina}, {Rodney}, {Rosario}, {Rosati}, {Salimbeni}, {Scarlata},
  {Siana}, {Simard}, {Smidt}, {Somerville}, {Spinrad}, {Straughn}, {Strolger},
  {Telford}, {Teplitz}, {Trump}, {van der Wel}, {Villforth}, {Wechsler},
  {Weiner}, {Wiklind}, {Wild}, {Wilson}, {Wuyts}, {Yan}, \& {Yun}}]{gro11}
{Grogin}, N.~A., {Kocevski}, D.~D., {Faber}, S.~M., {et~al.} 2011, \apjs, 197,
  35

\bibitem[{{Hainline} {et~al.}(2009){Hainline}, {Shapley}, {Kornei}, {Pettini},
  {Buckley-Geer}, {Allam}, \& {Tucker}}]{hai09}
{Hainline}, K.~N., {Shapley}, A.~E., {Kornei}, K.~A., {et~al.} 2009, \apj, 701,
  52

\bibitem[{{Henry} {et~al.}(2013){Henry}, {Scarlata}, {Dom{\'{\i}}nguez},
  {Malkan}, {Martin}, {Siana}, {Atek}, {Bedregal}, {Colbert}, {Rafelski},
  {Ross}, {Teplitz}, {Bunker}, {Dressler}, {Hathi}, {Masters}, {McCarthy}, \&
  {Straughn}}]{hen13}
{Henry}, A., {Scarlata}, C., {Dom{\'{\i}}nguez}, A., {et~al.} 2013, \apjl, 776,
  L27

\bibitem[{{Horne}(1986)}]{hor86}
{Horne}, K. 1986, \pasp, 98, 609

\bibitem[{{Jones} {et~al.}(2010){Jones}, {Ellis}, {Jullo}, \&
  {Richard}}]{jon10}
{Jones}, T., {Ellis}, R., {Jullo}, E., \& {Richard}, J. 2010, \apjl, 725, L176

\bibitem[{{Jones} {et~al.}(2013){Jones}, {Ellis}, {Richard}, \&
  {Jullo}}]{jon13}
{Jones}, T., {Ellis}, R.~S., {Richard}, J., \& {Jullo}, E. 2013, \apj, 765, 48

\bibitem[{{Kauffmann} {et~al.}(2003{\natexlab{a}}){Kauffmann}, {Heckman},
  {White}, {Charlot}, {Tremonti}, {Brinchmann}, {Bruzual}, {Peng}, {Seibert},
  {Bernardi}, {Blanton}, {Brinkmann}, {Castander}, {Cs{\'a}bai}, {Fukugita},
  {Ivezic}, {Munn}, {Nichol}, {Padmanabhan}, {Thakar}, {Weinberg}, \&
  {York}}]{kau03a}
{Kauffmann}, G., {Heckman}, T.~M., {White}, S.~D.~M., {et~al.}
  2003{\natexlab{a}}, \mnras, 341, 33

\bibitem[{{Kauffmann} {et~al.}(2003{\natexlab{b}}){Kauffmann}, {Heckman},
  {Tremonti}, {Brinchmann}, {Charlot}, {White}, {Ridgway}, {Brinkmann},
  {Fukugita}, {Hall}, {Ivezi{\'c}}, {Richards}, \& {Schneider}}]{kau03b}
{Kauffmann}, G., {Heckman}, T.~M., {Tremonti}, C., {et~al.} 2003{\natexlab{b}},
  \mnras, 346, 1055

\bibitem[{{Kennicutt}(1998)}]{ken98}
{Kennicutt}, Jr., R.~C. 1998, \araa, 36, 189

\bibitem[{{Kewley} \& {Dopita}(2002)}]{kew02}
{Kewley}, L.~J., \& {Dopita}, M.~A. 2002, \apjs, 142, 35

\bibitem[{{Kewley} {et~al.}(2013{\natexlab{a}}){Kewley}, {Dopita}, {Leitherer},
  {Dav{\'e}}, {Yuan}, {Allen}, {Groves}, \& {Sutherland}}]{kew13t}
{Kewley}, L.~J., {Dopita}, M.~A., {Leitherer}, C., {et~al.} 2013{\natexlab{a}},
  \apj, 774, 100

\bibitem[{{Kewley} {et~al.}(2001){Kewley}, {Dopita}, {Sutherland}, {Heisler},
  \& {Trevena}}]{kew01}
{Kewley}, L.~J., {Dopita}, M.~A., {Sutherland}, R.~S., {Heisler}, C.~A., \&
  {Trevena}, J. 2001, \apj, 556, 121

\bibitem[{{Kewley} \& {Ellison}(2008)}]{kew08}
{Kewley}, L.~J., \& {Ellison}, S.~L. 2008, \apj, 681, 1183

\bibitem[{{Kewley} {et~al.}(2013{\natexlab{b}}){Kewley}, {Maier}, {Yabe},
  {Ohta}, {Akiyama}, {Dopita}, \& {Yuan}}]{kew13}
{Kewley}, L.~J., {Maier}, C., {Yabe}, K., {et~al.} 2013{\natexlab{b}}, \apjl,
  774, L10

\bibitem[{{Kobulnicky} \& {Kewley}(2004)}]{kob04}
{Kobulnicky}, H.~A., \& {Kewley}, L.~J. 2004, \apj, 617, 240

\bibitem[{{Koekemoer} {et~al.}(2011){Koekemoer}, {Faber}, {Ferguson}, {Grogin},
  {Kocevski}, {Koo}, {Lai}, {Lotz}, {Lucas}, {McGrath}, {Ogaz}, {Rajan},
  {Riess}, {Rodney}, {Strolger}, {Casertano}, {Castellano}, {Dahlen},
  {Dickinson}, {Dolch}, {Fontana}, {Giavalisco}, {Grazian}, {Guo}, {Hathi},
  {Huang}, {van der Wel}, {Yan}, {Acquaviva}, {Alexander}, {Almaini}, {Ashby},
  {Barden}, {Bell}, {Bournaud}, {Brown}, {Caputi}, {Cassata}, {Challis},
  {Chary}, {Cheung}, {Cirasuolo}, {Conselice}, {Roshan Cooray}, {Croton},
  {Daddi}, {Dav{\'e}}, {de Mello}, {de Ravel}, {Dekel}, {Donley}, {Dunlop},
  {Dutton}, {Elbaz}, {Fazio}, {Filippenko}, {Finkelstein}, {Frazer}, {Gardner},
  {Garnavich}, {Gawiser}, {Gruetzbauch}, {Hartley}, {H{\"a}ussler},
  {Herrington}, {Hopkins}, {Huang}, {Jha}, {Johnson}, {Kartaltepe},
  {Khostovan}, {Kirshner}, {Lani}, {Lee}, {Li}, {Madau}, {McCarthy},
  {McIntosh}, {McLure}, {McPartland}, {Mobasher}, {Moreira}, {Mortlock},
  {Moustakas}, {Mozena}, {Nandra}, {Newman}, {Nielsen}, {Niemi}, {Noeske},
  {Papovich}, {Pentericci}, {Pope}, {Primack}, {Ravindranath}, {Reddy},
  {Renzini}, {Rix}, {Robaina}, {Rosario}, {Rosati}, {Salimbeni}, {Scarlata},
  {Siana}, {Simard}, {Smidt}, {Snyder}, {Somerville}, {Spinrad}, {Straughn},
  {Telford}, {Teplitz}, {Trump}, {Vargas}, {Villforth}, {Wagner}, {Wandro},
  {Wechsler}, {Weiner}, {Wiklind}, {Wild}, {Wilson}, {Wuyts}, \& {Yun}}]{koe11}
{Koekemoer}, A.~M., {Faber}, S.~M., {Ferguson}, H.~C., {et~al.} 2011, \apjs,
  197, 36

\bibitem[{{Kriek} {et~al.}(2009){Kriek}, {van Dokkum}, {Labb{\'e}}, {Franx},
  {Illingworth}, {Marchesini}, \& {Quadri}}]{kre09}
{Kriek}, M., {van Dokkum}, P.~G., {Labb{\'e}}, I., {et~al.} 2009, \apj, 700,
  221

\bibitem[{{Lara-L{\'o}pez} {et~al.}(2013){Lara-L{\'o}pez}, {Hopkins},
  {L{\'o}pez-S{\'a}nchez}, {Brough}, {Gunawardhana}, {Colless}, {Robotham},
  {Bauer}, {Bland-Hawthorn}, {Cluver}, {Driver}, {Foster}, {Kelvin}, {Liske},
  {Loveday}, {Owers}, {Ponman}, {Sharp}, {Steele}, {Taylor}, \&
  {Thomas}}]{lar13}
{Lara-L{\'o}pez}, M.~A., {Hopkins}, A.~M., {L{\'o}pez-S{\'a}nchez}, A.~R.,
  {et~al.} 2013, \mnras, 434, 451

\bibitem[{{Lilly} {et~al.}(2013){Lilly}, {Carollo}, {Pipino}, {Renzini}, \&
  {Peng}}]{lil13}
{Lilly}, S.~J., {Carollo}, C.~M., {Pipino}, A., {Renzini}, A., \& {Peng}, Y.
  2013, \apj, 772, 119

\bibitem[{{Liu} {et~al.}(2008){Liu}, {Shapley}, {Coil}, {Brinchmann}, \&
  {Ma}}]{liu08}
{Liu}, X., {Shapley}, A.~E., {Coil}, A.~L., {Brinchmann}, J., \& {Ma}, C.-P.
  2008, \apj, 678, 758

\bibitem[{{Maier} {et~al.}(2014){Maier}, {Lilly}, {Ziegler}, {Contini},
  {P{\'e}rez Montero}, {Peng}, \& {Balestra}}]{mai14}
{Maier}, C., {Lilly}, S.~J., {Ziegler}, B.~L., {et~al.} 2014, \apj, 792, 3

\bibitem[{{Maiolino} {et~al.}(2008){Maiolino}, {Nagao}, {Grazian}, {Cocchia},
  {Marconi}, {Mannucci}, {Cimatti}, {Pipino}, {Ballero}, {Calura}, {Chiappini},
  {Fontana}, {Granato}, {Matteucci}, {Pastorini}, {Pentericci}, {Risaliti},
  {Salvati}, \& {Silva}}]{mai08}
{Maiolino}, R., {Nagao}, T., {Grazian}, A., {et~al.} 2008, \aap, 488, 463

\bibitem[{{Mannucci} {et~al.}(2010){Mannucci}, {Cresci}, {Maiolino}, {Marconi},
  \& {Gnerucci}}]{man10}
{Mannucci}, F., {Cresci}, G., {Maiolino}, R., {Marconi}, A., \& {Gnerucci}, A.
  2010, \mnras, 408, 2115

\bibitem[{{Maseda} {et~al.}(2014){Maseda}, {van der Wel}, {Rix}, {da Cunha},
  {Pacifici}, {Momcheva}, {Brammer}, {Meidt}, {Franx}, {van Dokkum},
  {Fumagalli}, {Bell}, {Ferguson}, {F{\"o}rster-Schreiber}, {Koekemoer}, {Koo},
  {Lundgren}, {Marchesini}, {Nelson}, {Patel}, {Skelton}, {Straughn}, {Trump},
  \& {Whitaker}}]{mas14}
{Maseda}, M.~V., {van der Wel}, A., {Rix}, H.-W., {et~al.} 2014, ArXiv
  e-prints, arXiv:1406.3351

\bibitem[{{Masters} {et~al.}(2014){Masters}, {McCarthy}, {Siana}, {Malkan},
  {Mobasher}, {Atek}, {Henry}, {Martin}, {Rafelski}, {Hathi}, {Scarlata},
  {Ross}, {Bunker}, {Blanc}, {Bedregal}, {Dom{\'{\i}}nguez}, {Colbert},
  {Teplitz}, \& {Dressler}}]{mast14}
{Masters}, D., {McCarthy}, P., {Siana}, B., {et~al.} 2014, \apj, 785, 153

\bibitem[{{McLean} {et~al.}(2012){McLean}, {Steidel}, {Epps}, {Konidaris},
  {Matthews}, {Adkins}, {Aliado}, {Brims}, {Canfield}, {Cromer}, {Fucik},
  {Kulas}, {Mace}, {Magnone}, {Rodriguez}, {Rudie}, {Trainor}, {Wang}, {Weber},
  \& {Weiss}}]{mcl12}
{McLean}, I.~S., {Steidel}, C.~C., {Epps}, H.~W., {et~al.} 2012, in Society of
  Photo-Optical Instrumentation Engineers (SPIE) Conference Series, Vol. 8446,
  Society of Photo-Optical Instrumentation Engineers (SPIE) Conference Series

\bibitem[{{Newman} {et~al.}(2014){Newman}, {Buschkamp}, {Genzel}, {F{\"o}rster
  Schreiber}, {Kurk}, {Sternberg}, {Gnat}, {Rosario}, {Mancini}, {Lilly},
  {Renzini}, {Burkert}, {Carollo}, {Cresci}, {Davies}, {Eisenhauer}, {Genel},
  {Shapiro Griffin}, {Hicks}, {Lutz}, {Naab}, {Peng}, {Tacconi}, {Wuyts},
  {Zamorani}, {Vergani}, \& {Weiner}}]{new14}
{Newman}, S.~F., {Buschkamp}, P., {Genzel}, R., {et~al.} 2014, \apj, 781, 21

\bibitem[{{Papovich} {et~al.}(2011){Papovich}, {Finkelstein}, {Ferguson},
  {Lotz}, \& {Giavalisco}}]{pap11}
{Papovich}, C., {Finkelstein}, S.~L., {Ferguson}, H.~C., {Lotz}, J.~M., \&
  {Giavalisco}, M. 2011, \mnras, 412, 1123

\bibitem[{{Pettini} \& {Pagel}(2004)}]{pet04}
{Pettini}, M., \& {Pagel}, B.~E.~J. 2004, \mnras, 348, L59

\bibitem[{{Pilyugin} \& {Thuan}(2005)}]{pil05}
{Pilyugin}, L.~S., \& {Thuan}, T.~X. 2005, \apj, 631, 231

\bibitem[{{Queyrel} {et~al.}(2012){Queyrel}, {Contini}, {Kissler-Patig},
  {Epinat}, {Amram}, {Garilli}, {Le F{\`e}vre}, {Moultaka}, {Paioro}, {Tasca},
  {Tresse}, {Vergani}, {L{\'o}pez-Sanjuan}, \& {Perez-Montero}}]{que12}
{Queyrel}, J., {Contini}, T., {Kissler-Patig}, M., {et~al.} 2012, \aap, 539,
  A93

\bibitem[{{Reddy} {et~al.}(2012){Reddy}, {Pettini}, {Steidel}, {Shapley},
  {Erb}, \& {Law}}]{red12}
{Reddy}, N.~A., {Pettini}, M., {Steidel}, C.~C., {et~al.} 2012, \apj, 754, 25

\bibitem[{{Rigby} {et~al.}(2011){Rigby}, {Wuyts}, {Gladders}, {Sharon}, \&
  {Becker}}]{rig11}
{Rigby}, J.~R., {Wuyts}, E., {Gladders}, M.~D., {Sharon}, K., \& {Becker},
  G.~D. 2011, \apj, 732, 59

\bibitem[{{Shapley} {et~al.}(2005){Shapley}, {Coil}, {Ma}, \& {Bundy}}]{sha05}
{Shapley}, A.~E., {Coil}, A.~L., {Ma}, C.-P., \& {Bundy}, K. 2005, \apj, 635,
  1006

\bibitem[{{Shapley} {et~al.}(2014){Shapley}, {Reddy}, {Kriek}, {Freeman},
  {Sanders}, {Siana}, {Coil}, {Mobasher}, {Shivaei}, {Price}, \& {de
  Groot}}]{sha14}
{Shapley}, A.~E., {Reddy}, N.~A., {Kriek}, M., {et~al.} 2014, ArXiv e-prints,
  arXiv:1409.7071

\bibitem[{{Skelton} {et~al.}(2014){Skelton}, {Whitaker}, {Momcheva}, {Brammer},
  {van Dokkum}, {Labbe}, {Franx}, {van der Wel}, {Bezanson}, {Da Cunha},
  {Fumagalli}, {Foerster Schreiber}, {Kriek}, {Leja}, {Lundgren}, {Magee},
  {Marchesini}, {Maseda}, {Nelson}, {Oesch}, {Pacifici}, {Patel}, {Price},
  {Rix}, {Tal}, {Wake}, \& {Wuyts}}]{ske14}
{Skelton}, R.~E., {Whitaker}, K.~E., {Momcheva}, I.~G., {et~al.} 2014, ArXiv
  e-prints, arXiv:1403.3689

\bibitem[{{Steidel} {et~al.}(2014){Steidel}, {Rudie}, {Strom}, {Pettini},
  {Reddy}, {Shapley}, {Trainor}, {Erb}, {Turner}, {Konidaris}, {Kulas}, {Mace},
  {Matthews}, \& {McLean}}]{ste14}
{Steidel}, C.~C., {Rudie}, G.~C., {Strom}, A.~L., {et~al.} 2014, ArXiv
  e-prints, arXiv:1405.5473

\bibitem[{{Stott} {et~al.}(2013){Stott}, {Sobral}, {Bower}, {Smail}, {Best},
  {Matsuda}, {Hayashi}, {Geach}, \& {Kodama}}]{sto13}
{Stott}, J.~P., {Sobral}, D., {Bower}, R., {et~al.} 2013, \mnras, 436, 1130

\bibitem[{{Stott} {et~al.}(2014){Stott}, {Sobral}, {Swinbank}, {Smail},
  {Bower}, {Best}, {Sharples}, {Geach}, \& {Matthee}}]{sto14}
{Stott}, J.~P., {Sobral}, D., {Swinbank}, A.~M., {et~al.} 2014, \mnras, 443,
  2695

\bibitem[{{Tacconi} {et~al.}(2010){Tacconi}, {Genzel}, {Neri}, {Cox}, {Cooper},
  {Shapiro}, {Bolatto}, {Bouch{\'e}}, {Bournaud}, {Burkert}, {Combes},
  {Comerford}, {Davis}, {Schreiber}, {Garcia-Burillo}, {Gracia-Carpio}, {Lutz},
  {Naab}, {Omont}, {Shapley}, {Sternberg}, \& {Weiner}}]{tac10}
{Tacconi}, L.~J., {Genzel}, R., {Neri}, R., {et~al.} 2010, \nat, 463, 781

\bibitem[{{Tacconi} {et~al.}(2013){Tacconi}, {Neri}, {Genzel}, {Combes},
  {Bolatto}, {Cooper}, {Wuyts}, {Bournaud}, {Burkert}, {Comerford}, {Cox},
  {Davis}, {F{\"o}rster Schreiber}, {Garc{\'{\i}}a-Burillo}, {Gracia-Carpio},
  {Lutz}, {Naab}, {Newman}, {Omont}, {Saintonge}, {Shapiro Griffin}, {Shapley},
  {Sternberg}, \& {Weiner}}]{tac13}
{Tacconi}, L.~J., {Neri}, R., {Genzel}, R., {et~al.} 2013, \apj, 768, 74

\bibitem[{{Tassis} {et~al.}(2008){Tassis}, {Kravtsov}, \& {Gnedin}}]{tas08}
{Tassis}, K., {Kravtsov}, A.~V., \& {Gnedin}, N.~Y. 2008, \apj, 672, 888

\bibitem[{{Tremonti} {et~al.}(2004){Tremonti}, {Heckman}, {Kauffmann},
  {Brinchmann}, {Charlot}, {White}, {Seibert}, {Peng}, {Schlegel}, {Uomoto},
  {Fukugita}, \& {Brinkmann}}]{tre04}
{Tremonti}, C.~A., {Heckman}, T.~M., {Kauffmann}, G., {et~al.} 2004, \apj, 613,
  898

\bibitem[{{Troncoso} {et~al.}(2014){Troncoso}, {Maiolino}, {Sommariva},
  {Cresci}, {Mannucci}, {Marconi}, {Meneghetti}, {Grazian}, {Cimatti},
  {Fontana}, {Nagao}, \& {Pentericci}}]{tro14}
{Troncoso}, P., {Maiolino}, R., {Sommariva}, V., {et~al.} 2014, \aap, 563, A58

\bibitem[{{Vila-Costas} \& {Edmunds}(1992)}]{vil92}
{Vila-Costas}, M.~B., \& {Edmunds}, M.~G. 1992, \mnras, 259, 121

\bibitem[{{Whitaker} {et~al.}(2014){Whitaker}, {Franx}, {Leja}, {van Dokkum},
  {Henry}, {Skelton}, {Fumagalli}, {Momcheva}, {Brammer}, {Labbe}, {Nelson}, \&
  {Rigby}}]{whi14}
{Whitaker}, K.~E., {Franx}, M., {Leja}, J., {et~al.} 2014, ArXiv e-prints,
  arXiv:1407.1843

\bibitem[{{Wuyts} {et~al.}(2014){Wuyts}, {Kurk}, {F{\"o}rster Schreiber},
  {Genzel}, {Wisnioski}, {Bandara}, {Wuyts}, {Beifiori}, {Bender}, {Brammer},
  {Burkert}, {Buschkamp}, {Carollo}, {Chan}, {Davies}, {Eisenhauer}, {Fossati},
  {Kulkarni}, {Lang}, {Lilly}, {Lutz}, {Mancini}, {Mendel}, {Momcheva}, {Naab},
  {Nelson}, {Renzini}, {Rosario}, {Saglia}, {Seitz}, {Sharples}, {Sternberg},
  {Tacchella}, {Tacconi}, {van Dokkum}, \& {Wilman}}]{wuy14}
{Wuyts}, E., {Kurk}, J., {F{\"o}rster Schreiber}, N.~M., {et~al.} 2014, \apjl,
  789, L40

\bibitem[{{York} {et~al.}(2000){York}, {Adelman}, {Anderson}, {Anderson},
  {Annis}, {Bahcall}, {Bakken}, {Barkhouser}, {Bastian}, {Berman}, {Boroski},
  {Bracker}, {Briegel}, {Briggs}, {Brinkmann}, {Brunner}, {Burles}, {Carey},
  {Carr}, {Castander}, {Chen}, {Colestock}, {Connolly}, {Crocker}, {Csabai},
  {Czarapata}, {Davis}, {Doi}, {Dombeck}, {Eisenstein}, {Ellman}, {Elms},
  {Evans}, {Fan}, {Federwitz}, {Fiscelli}, {Friedman}, {Frieman}, {Fukugita},
  {Gillespie}, {Gunn}, {Gurbani}, {de Haas}, {Haldeman}, {Harris}, {Hayes},
  {Heckman}, {Hennessy}, {Hindsley}, {Holm}, {Holmgren}, {Huang}, {Hull},
  {Husby}, {Ichikawa}, {Ichikawa}, {Ivezi{\'c}}, {Kent}, {Kim}, {Kinney},
  {Klaene}, {Kleinman}, {Kleinman}, {Knapp}, {Korienek}, {Kron}, {Kunszt},
  {Lamb}, {Lee}, {Leger}, {Limmongkol}, {Lindenmeyer}, {Long}, {Loomis},
  {Loveday}, {Lucinio}, {Lupton}, {MacKinnon}, {Mannery}, {Mantsch}, {Margon},
  {McGehee}, {McKay}, {Meiksin}, {Merelli}, {Monet}, {Munn}, {Narayanan},
  {Nash}, {Neilsen}, {Neswold}, {Newberg}, {Nichol}, {Nicinski}, {Nonino},
  {Okada}, {Okamura}, {Ostriker}, {Owen}, {Pauls}, {Peoples}, {Peterson},
  {Petravick}, {Pier}, {Pope}, {Pordes}, {Prosapio}, {Rechenmacher}, {Quinn},
  {Richards}, {Richmond}, {Rivetta}, {Rockosi}, {Ruthmansdorfer}, {Sandford},
  {Schlegel}, {Schneider}, {Sekiguchi}, {Sergey}, {Shimasaku}, {Siegmund},
  {Smee}, {Smith}, {Snedden}, {Stone}, {Stoughton}, {Strauss}, {Stubbs},
  {SubbaRao}, {Szalay}, {Szapudi}, {Szokoly}, {Thakar}, {Tremonti}, {Tucker},
  {Uomoto}, {Vanden Berk}, {Vogeley}, {Waddell}, {Wang}, {Watanabe},
  {Weinberg}, {Yanny}, {Yasuda}, \& {SDSS Collaboration}}]{yor00}
{York}, D.~G., {Adelman}, J., {Anderson}, Jr., J.~E., {et~al.} 2000, \aj, 120,
  1579

\bibitem[{{Yuan} \& {Kewley}(2009)}]{yua09}
{Yuan}, T.-T., \& {Kewley}, L.~J. 2009, \apjl, 699, L161

\bibitem[{{Zahid} {et~al.}(2013){Zahid}, {Kashino}, {Silverman}, {Kewley},
  {Daddi}, {Renzini}, {Rodighiero}, {Nagao}, {Arimoto}, {Sanders},
  {Kartaltepe}, {Lilly}, {Maier}, {Geller}, {Capak}, {Carollo}, {Chu},
  {Hasinger}, {Ilbert}, {Kajisawa}, {Koekemoer}, {Kovac}, {Le Fevre},
  {Masters}, {McCracken}, {Onodera}, {Scoville}, {Strazzullo}, {Sugiyama},
  {Taniguchi}, \& {The COSMOS Team}}]{zah13}
{Zahid}, H.~J., {Kashino}, D., {Silverman}, J.~D., {et~al.} 2013, ArXiv
  e-prints, arXiv:1310.4950

\bibitem[{{Zahid} {et~al.}(2014){Zahid}, {Dima}, {Kudritzki}, {Kewley},
  {Geller}, {Hwang}, {Silverman}, \& {Kashino}}]{zah14}
{Zahid}, J., {Dima}, G., {Kudritzki}, R., {et~al.} 2014, ArXiv e-prints,
  arXiv:1404.7526

\bibitem[{{Zaritsky} {et~al.}(1994){Zaritsky}, {Kennicutt}, \&
  {Huchra}}]{zar94}
{Zaritsky}, D., {Kennicutt}, Jr., R.~C., \& {Huchra}, J.~P. 1994, \apj, 420, 87

\end{thebibliography}

\end{document}